\documentclass[pdftex,twocolumn,epjc3_preprint,runningheads]{svjour3}

\usepackage[colorlinks,citecolor=blue,urlcolor=blue,linkcolor=blue,breaklinks=true]{hyperref}
\usepackage[dvipsnames]{xcolor}

\pdfoutput=1

\usepackage[T1]{fontenc}
\usepackage{lmodern}
\usepackage{calc}
\usepackage{graphicx}
\usepackage{booktabs}
\usepackage{textcomp}
\usepackage{xspace}
\usepackage{relsize}
\usepackage{amssymb}
\usepackage{amsmath}
\usepackage{listings}
\usepackage{microtype}
\usepackage{multirow}
\usepackage{tabularx}
\usepackage{array}
\usepackage{placeins}
\usepackage{cuted}
\usepackage{soul} 
\usepackage{fixltx2e}
\usepackage{slashed}
\usepackage{bm}
\usepackage[numbers,sort&compress]{natbib}
\usepackage[labelfont=bf,font=small]{caption}
\usepackage[skip=-2pt]{subcaption}
\usepackage[clockwise,figuresright]{rotating}
\usepackage{tikz}
\usepackage[normalem]{ulem}
\usepackage[utf8]{inputenc}

\usepackage{etoolbox}
\AfterEndEnvironment{strip}{\leavevmode}

\allowdisplaybreaks

\newcolumntype{L}{>{\raggedright\let\newline\\\arraybackslash\hspace{0pt}}X}
\newcolumntype{R}{>{\raggedleft\let\newline\\\arraybackslash\hspace{0pt}}X}
\newcolumntype{C}{>{\centering\let\newline\\\arraybackslash\hspace{0pt}}X}

\newcommand{\gambitinstitute}[1]{\expandafter\csname #1\endcsname\label{#1}}
\newcommand{\gi}[1]{\gambitinstitute{#1}\and}
\newcommand{\last}[1]{\gambitinstitute{#1}}

\makeatletter

\newcommand{\preprintnumber}[1]{\gdef\@preprintnumber{\begin{flushright}{#1}\end{flushright}}}

\g@addto@macro\bfseries{\boldmath}
\makeatother

\let\underscore\_
\renewcommand{\_}{\discretionary{\underscore}{}{\underscore}}

\makeatletter
\let\orgdescriptionlabel\descriptionlabel
\renewcommand*{\descriptionlabel}[1]{%
  \let\orglabel\label
  \let\label\@gobble
  \phantomsection
  \protected@edef\@currentlabel{#1}%
  \let\label\orglabel
  \orgdescriptionlabel{#1}%
}
\makeatother

\newcommand\postnewlinemarker{\hbox{\ensuremath{\hookrightarrow}}}
\newcommand\cpp[1]{{\lstinline!#1!}}  

\newcommand\yaml[1]{{\lstset{style=yaml}\lstinline!#1!\lstset{style=cpp}}}
\newcommand\yamlvalue[1]{{\YAMLvaluestyle\ttfamily#1}}

\newcommand\term[1]{{\lstset{style=terminal}\lstinline!#1!\lstset{style=cpp}}}
\newcommand\termalt[1]{{\lstset{style=terminalalt}\lstinline!#1!\lstset{style=cpp}}}
\newcommand\fortran[1]{{\lstset{style=fortran}\lstinline!#1!\lstset{style=cpp}}}
\newcommand\py[1]{{\lstset{style=python}\lstinline!#1!\lstset{style=cpp}}}
\newcommand\customtilde{{\raisebox{0.2ex}{\scalebox{0.6}{\boldmath$\sim$}}}}
\newcommand\mathematica[1]{{\lstset{style=Mathematica}\lstinline!#1!\lstset{style=cpp}}}
\newcommand\guminline[1]{{{\lstset{style=gum}\lstinline!#1!}}}
\newcommand\textinline[1]{{{\lstset{style=text}\lstinline!#1!}}}

\def\be{\begin{equation}}
\def\ee{\end{equation}}
\def\ba{\begin{eqnarray}}
\def\ea{\end{eqnarray}}
\newcommand{\bea}{\begin{eqnarray}}
\newcommand{\eea}{\end{eqnarray}}

\lstnewenvironment{lstlistingyaml}{\lstset{style=yaml}}{\lstset{style=cpp}}
\lstnewenvironment{lstlistingterm}{\lstset{style=terminal}}{\lstset{style=cpp}}
\lstnewenvironment{lstlistingfortran}{\lstset{style=fortran}}{\lstset{style=cpp}}
\lstnewenvironment{lstcpp}{\lstset{style=cpp}}{\lstset{style=cpp}}
\lstnewenvironment{lstcppalt}{\lstset{style=cppalt}}{\lstset{style=cpp}}
\lstnewenvironment{lstcppnum}{\lstset{style=cppnum}}{\lstset{style=cpp}}
\lstnewenvironment{lstyaml}{\lstset{style=yaml}}{\lstset{style=cpp}}
\lstnewenvironment{lstgum}{\lstset{style=gum}}{\lstset{style=cpp}}
\lstnewenvironment{lstterm}{\lstset{style=terminal}}{\lstset{style=cpp}}
\lstnewenvironment{lsttermalt}{\lstset{style=terminalalt}}{\lstset{style=cpp}}
\lstnewenvironment{lsttext}{\lstset{style=text}}{\lstset{style=cpp}}
\lstnewenvironment{lstfortran}{\lstset{style=fortran}}{\lstset{style=cpp}}
\lstnewenvironment{lstpy}{\lstset{style=python}}{\lstset{style=cpp}}
\lstnewenvironment{lstmathematica}{\lstset{style=mathematica}}{\lstset{style=cpp}}

\newcommand{\tmpname}{}
\newcommand{\tmplistingname}{}
\makeatletter
\newif\ifATOlabelname
\lst@Key{labelname}{Listing}{\def\ATOlabelname{#1}\global\ATOlabelnametrue}
\makeatother
\lstnewenvironment{lstcpplabel}[1][]{
  \lstset{style=cpp,#1} 
  \ifATOlabelname
    \renewcommand{\tmpname}{\lstlistingname}
    \renewcommand{\tmplistingname}{\lstlistlistingname}
    \renewcommand{\lstlistingname}{\ATOlabelname}
    \renewcommand{\lstlistlistingname}{List of \lstlistingname s}
  \fi
}{
  \renewcommand{\lstlistingname}{\tmpname}
  \renewcommand{\lstlistlistingname}{\tmplistingname}
  \lstset{style=cpp}
}
\definecolor{solarized@base03}{HTML}{002B36}
\definecolor{solarized@base02}{HTML}{073642}
\definecolor{solarized@base01}{HTML}{586e75}
\definecolor{solarized@base00}{HTML}{657b83}
\definecolor{solarized@base0}{HTML}{839496}
\definecolor{solarized@base1}{HTML}{93a1a1}
\definecolor{solarized@base2}{HTML}{EEE8D5}
\definecolor{solarized@base3}{HTML}{FDF6E3}
\definecolor{solarized@yellow}{HTML}{B58900}
\definecolor{solarized@orange}{HTML}{CB4B16}
\definecolor{solarized@red}{HTML}{DC322F}
\definecolor{solarized@magenta}{HTML}{D33682}
\definecolor{solarized@violet}{HTML}{6C71C4}
\definecolor{solarized@blue}{HTML}{268BD2}
\definecolor{solarized@cyan}{HTML}{2AA198}
\definecolor{solarized@green}{HTML}{859900}
\definecolor{darkred}{HTML}{550003}
\definecolor{darkgreen}{HTML}{00AA00}
\definecolor{orchid}{HTML}{AF06F5}

\newcommand\YAMLstringstyle{\footnotesize\color{solarized@green}\mdseries}
\newcommand\YAMLkeystyle{\footnotesize\color{solarized@blue}\ttfamily}
\newcommand\YAMLvaluestyle{\footnotesize\color{blue}\mdseries}
\newcommand\ProcessThreeDashes{\llap{\color{cyan}\mdseries-{-}-}}

\newcommand\CPPcommentstyle{\color{solarized@violet}\footnotesize\ttfamily}
\newcommand\CPPdirectivestyle{\color{solarized@magenta}\footnotesize\ttfamily}
\newcommand\termplainstyle{\footnotesize\ttfamily}

\newcommand\YAMLcommentstyle{\color{solarized@orange}\ttfamily}

\newcommand\processLongMacroDelimiter
{%
\CPPdirectivestyle%
\#define%
}

\lstdefinestyle{cpp}
{
  language=C++,
  basicstyle=\footnotesize\ttfamily,
  basewidth={0.53em,0.44em}, 
  numbers=none,
  tabsize=2,
  breaklines=true,
  escapeinside={@}{@},
  showstringspaces=false,
  numberstyle=\tiny\color{solarized@base01},
  keywordstyle=\color{solarized@orange},
  stringstyle=\color{solarized@red}\ttfamily,
  identifierstyle=\color{solarized@blue},
  commentstyle=\CPPcommentstyle,
  directivestyle=\CPPdirectivestyle,
  emphstyle=\color{solarized@green},
  frame=single,
  rulecolor=\color{solarized@base2},
  rulesepcolor=\color{solarized@base2},
  literate={~} {\customtilde}1,
  moredelim=*[directive]\ \ \#,
  moredelim=*[directive]\ \ \ \ \#
}

\lstdefinestyle{cppalt}
{
  language=C++,
  basicstyle=\footnotesize\ttfamily,
  basewidth={0.53em,0.44em}, 
  numbers=none,
  tabsize=2,
  breaklines=true,
  escapeinside={*@}{@*},
  showstringspaces=false,
  numberstyle=\tiny\color{solarized@base01},
  keywordstyle=\color{solarized@orange},
  stringstyle=\color{solarized@red}\ttfamily,
  identifierstyle=\color{solarized@blue},
  commentstyle=\CPPcommentstyle,
  directivestyle=\CPPdirectivestyle,
  emphstyle=\color{solarized@green},
  frame=single,
  rulecolor=\color{solarized@base2},
  rulesepcolor=\color{solarized@base2},
  literate={~}{\customtilde}1,
  moredelim=**[is][\processLongMacroDelimiter]{BeginLongMacro}{EndLongMacro} 
}

\lstdefinestyle{cppnum}
{
  language=C++,
  basicstyle=\footnotesize\ttfamily,
  basewidth={0.53em,0.44em}, 
  numbers=none,
  tabsize=2,
  breaklines=true,
  escapeinside={@}{@},
  numberstyle=\tiny\color{solarized@base01},
  showstringspaces=false,
  keywordstyle=\color{solarized@orange},
  stringstyle=\color{solarized@red}\ttfamily,
  identifierstyle=\color{solarized@blue},
  commentstyle=\CPPcommentstyle,
  directivestyle=\CPPdirectivestyle,
  emphstyle=\color{solarized@green},
  frame=single,
  rulecolor=\color{solarized@base2},
  rulesepcolor=\color{solarized@base2},
  literate={~} {\customtilde}1,
  moredelim=*[directive]\ \ \#,
  moredelim=*[directive]\ \ \ \ \#
}

\lstdefinestyle{python}
{
  language=Python,
  basicstyle=\footnotesize\ttfamily,
  basewidth={0.53em,0.44em},
  numbers=none,
  tabsize=2,
  breaklines=true,
  escapeinside={@}{@},
  showstringspaces=false,
  numberstyle=\tiny\color{solarized@base01},
  keywordstyle=\color{blue},
  stringstyle=\color{orange}\ttfamily,
  identifierstyle=\color{darkred},
  commentstyle=\color{purple},
  emphstyle=\color{green},
  frame=single,
  rulecolor=\color{solarized@base2},
  rulesepcolor=\color{solarized@base2},
  literate = {~}{\customtilde}1
             {\ as\ }{{\color{blue}\ as\ \color{black}}}3
             {.set}{{\color{black}.}{\color{darkred}set}}4
}

\lstdefinestyle{fortran}
{
  language=Fortran,
  basicstyle=\footnotesize\ttfamily,
  basewidth={0.53em,0.44em},
  numbers=none,
  tabsize=2,
  breaklines=true,
  escapeinside={@}{@},
  showstringspaces=false,
  numberstyle=\tiny\color{solarized@base01},
  keywordstyle=\color{blue},
  stringstyle=\color{orange}\ttfamily,
  identifierstyle=\color{Periwinkle},
  commentstyle=\color{purple},
  emphstyle=\color{green},
  morekeywords={and, or, true, false},
  frame=single,
  rulecolor=\color{solarized@base2},
  rulesepcolor=\color{solarized@base2},
  literate={~}{\customtilde}1
}

\lstdefinestyle{terminal}
{
  language=bash,
  basicstyle=\termplainstyle,
  numbers=none,
  tabsize=2,
  breaklines=true,
  escapeinside={@}{@},
  frame=single,
  showstringspaces=false,
  numberstyle=\tiny\color{solarized@base01},
  keywordstyle=\color{solarized@orange},
  stringstyle=\color{solarized@red}\ttfamily,
  identifierstyle=\color{black},
  commentstyle=\color{solarized@violet},
  emphstyle=\color{solarized@green},
  frame=single,
  rulecolor=\color{solarized@base2},
  rulesepcolor=\color{solarized@base2},
  morekeywords={gambit, cmake, make, mkdir, gum, python, wget, tar, cp, pippi, mpirun},
  deletekeywords={test},
  literate = {/gambit}{{/}{\color{black}}gambit}6
             {gambit/}{{\color{black}}gambit{/}}6
             {gum/}{{\color{black}}gum{/}}4
             {/include}{{/}{\color{black}}include}8
             {cmake/}{{\color{black}}cmake/}6
             {.cmake}{{.}{\color{black}}cmake}6
             {.gum}{{.}{\color{black}}gum}6
             {.tar}{{.}{\color{black}}tar}4
             {source/}{{\color{black}}source{/}}7
             { type}{{\color{black}}{}type}5
             {~}{\customtilde}1
             {math}{{\color{solarized@orange}}math}4
}

\lstdefinestyle{terminalalt}
{
  language=bash,
  basicstyle=\footnotesize\ttfamily,
  numbers=none,
  tabsize=2,
  breaklines=true,
  escapeinside={*@}{@*},
  frame=single,
  showstringspaces=false,
  numberstyle=\tiny\color{solarized@base01},
  keywordstyle=\color{solarized@orange},
  stringstyle=\color{solarized@red}\ttfamily,
  identifierstyle=\color{black},
  commentstyle=\color{solarized@violet},
  emphstyle=\color{solarized@green},
  frame=single,
  rulecolor=\color{solarized@base2},
  rulesepcolor=\color{solarized@base2},
  morekeywords={gambit, cmake, make, mkdir},
  deletekeywords={test},
  literate = {\ gambit}{{\ }{\color{black}}gambit}7
             {/gambit}{{/}{\color{black}}gambit}6
             {gambit/}{{\color{black}}gambit{/}}6
             {/include}{{/}{\color{black}}include}8
             {cmake/}{{\color{black}}cmake/}6
             {.cmake}{{.}{\color{black}}cmake}6
             {~}{\customtilde}1
}

\lstdefinestyle{text}
{
  language={},
  basicstyle=\footnotesize\ttfamily,
  identifierstyle=\color{black},
  numbers=none,
  tabsize=2,
  breaklines=true,
  escapeinside={*@}{@*},
  showstringspaces=false,
  frame=single,
  rulecolor=\color{solarized@base2},
  rulesepcolor=\color{solarized@base2},
  literate={~}{\customtilde}1
}

\lstdefinestyle{yaml}
{
  language=bash,
  escapeinside={@}{@},
  keywords={true,false,null},
  otherkeywords={},
  keywordstyle=\color{solarized@base0}\bfseries,
  basicstyle=\footnotesize\color{black}\ttfamily,
  identifierstyle=\YAMLkeystyle,
  sensitive=false,
  commentstyle=\YAMLcommentstyle,
  morecomment=[l]{\#},
  morecomment=[s]{/*}{*/},
  stringstyle=\YAMLstringstyle\ttfamily,
  moredelim=**[s][\YAMLkeystyle]{,}{:},   
  moredelim=**[l][\YAMLvaluestyle]{:},    
  morestring=[b]',
  morestring=[b]",
  literate =    {---}{{\ProcessThreeDashes}}3
                {>}{{\textcolor{solarized@red}\textgreater}}1
                {gtr}{\textgreater}1
                {grt}{\textgreater}1
                {|}{{\textcolor{solarized@red}\textbar}}1
                {\ -\ }{{\mdseries\color{black}\ -\ \negmedspace}}3
                {\}}{{{\color{black} \}}}}1
                {\{}{{{\color{black} \{}}}1
                {[}{{{\color{black} [}}}1
                {]}{{{\color{black} ]}}}1
                {~}{\customtilde}1,
  breakindent=0pt,
  breakatwhitespace,
  columns=fullflexible
}

\lstdefinestyle{gum}
{
  language=bash,
  escapeinside={@}{@},
  keywords={true,false,null,all},
  otherkeywords={},
  keywordstyle=\color{solarized@base02}\bfseries,
  basicstyle=\footnotesize\color{black}\ttfamily,
  identifierstyle=\color{solarized@magenta},
  sensitive=false,
  commentstyle=\color{solarized@cyan}\ttfamily,
  morecomment=[l]{\#},
  morecomment=[s]{/*}{*/},
  stringstyle=\footnotesize\color{solarized@base01}\mdseries\ttfamily,
  moredelim=**[l][\footnotesize\color{solarized@base02}\mdseries]{:},    
  morestring=[b]',
  morestring=[b]",
  literate =    {---}{{\ProcessThreeDashes}}3
                {grt}{{\textcolor{solarized@magenta}\textgreater}}1
                {gtr}{{\textcolor{solarized@base02}\textgreater}}1
                {/>}{{\textcolor{solarized@magenta}\textgreater}}1
                {/<}{{\textcolor{solarized@magenta}\textless}}1
                {lss}{{\textcolor{solarized@base02}\textless}}1
                {pls}{{\textcolor{solarized@magenta}+}}1
                {mns}{{\textcolor{solarized@magenta}-}}1
                {|}{{\textcolor{solarized@base02}\textbar}}1
                {\ -\ }{{\mdseries\color{solarized@base02}\ -\ \negmedspace}}3
                {\}}{{{\color{solarized@base02} \}}}}1
                {\{}{{{\color{solarized@base02} \{}}}1
                {[}{{{\color{solarized@base02} [}}}1
                {]}{{{\color{solarized@base02} ]}}}1
                {~}{\customtilde}1,
  breakindent=0pt,
  breakatwhitespace,
  columns=fullflexible
}

\lstdefinestyle{mathematica}
{
  language={Mathematica},
  basicstyle=\footnotesize\ttfamily,
  basewidth={0.53em,0.44em},
  numbers=none,
  tabsize=2,
  breaklines=true,
  postbreak=,
  escapeinside={@}{@},
  numberstyle=\tiny\color{black},
  showstringspaces=false,
  numberstyle=\tiny\color{solarized@base01},
  keywordstyle=\color{solarized@orange},
  stringstyle=\color{solarized@red}\ttfamily,
  identifierstyle=\color{solarized@orange}\ttfamily,
  commentstyle=\color{solarized@gray}\ttfamily,
  directivestyle=\color{solarized@orange}\ttfamily,
  emphstyle=\color{solarized@green},
  frame=single,
  rulecolor=\color{solarized@base2},
  rulesepcolor=\color{solarized@base2},
  literate={~} {\customtilde}1,
  moredelim=*[directive]\ \ \#,
  moredelim=*[directive]\ \ \ \ \#,
  mathescape=false
}

\newcommand{\doublecross}[2]{\hyperref[#2]{\textbf{#1}}}
\newcommand{\doublecrosssf}[2]{\hyperref[#2]{\textbf{\textsf{#1}}}}

\newcommand{\startglossary}{\section{Glossary}\label{glossary}Here we explain some terms that have specific technical definitions in \GB.\begin{description}}
\newcommand{\finishglossary}{\end{description}}

\newcommand{\bcode}{\begin{lstlisting}}
\newcommand{\ecode}{\end{lstlisting}}
\newcommand{\metavarf}[1]{\textit{\color{darkgreen}\footnotesize\textrm{#1}}}

\newcommand{\metavar}{\metavarf}

\newcommand{\ie}{i.e.\ }
\newcommand{\eg}{e.g.\ }

\newcommand{\gambit}{\textsf{GAMBIT}\xspace}

\newcommand{\darkbit}{\textsf{DarkBit}\xspace}

\newcommand{\colliderbit}{\textsf{ColliderBit}\xspace}
\newcommand{\flavbit}{\textsf{FlavBit}\xspace}
\newcommand{\specbit}{\textsf{SpecBit}\xspace}
\newcommand{\decaybit}{\textsf{DecayBit}\xspace}

\newcommand{\scannerbit}{\textsf{ScannerBit}\xspace}

\newcommand{\BOSS}{\textsf{BOSS}\xspace}
\newcommand{\GB}{\gambit}

\newcommand{\mpi}{\textsf{MPI}\xspace}

\newcommand{\pythia}{\textsf{Pythia}\xspace}

\newcommand{\madgraph}{\textsf{MadGraph}\xspace}

\newcommand{\higgsbounds}{\textsf{HiggsBounds}\xspace}

\newcommand{\higgssignals}{\textsf{HiggsSignals}\xspace}
\newcommand{\ds}{\textsf{DarkSUSY}\xspace}
\newcommand{\darksusy}{\ds}

\newcommand{\mo}{\micromegas}
\newcommand{\micromegas}{\textsf{micrOMEGAs}\xspace}

\newcommand\FS{\FlexibleSUSY}
\newcommand\flexiblesusy{\FlexibleSUSY}
\newcommand\FlexibleSUSY{\textsf{FlexibleSUSY}\xspace}

\newcommand\Mathematica{\textsf{Mathematica}\xspace}

\newcommand\nulike{\textsf{nulike}\xspace}
\newcommand\gamLike{\textsf{gamLike}\xspace}
\newcommand\gamlike{\gamLike}

\newcommand\pippi{\textsf{pippi}\xspace}

\newcommand\diver{\textsf{Diver}\xspace}
\newcommand\ddcalc{\textsf{DDCalc}\xspace}

\newcommand{\gum}{\textsf{GUM}\xspace}
\newcommand{\dgum}{\!\!\term{.gum}\!\xspace}
\newcommand{\fr}{\textsf{FeynRules}\xspace}
\newcommand{\sarah}{\textsf{SARAH}\xspace}
\newcommand{\CH}{\textsf{CalcHEP}\xspace}
\newcommand{\MG}{\textsf{MadGraph}\xspace}
\newcommand{\mdm}{\textsf{MadDM}\xspace}
\newcommand{\ufo}{\textsf{UFO}\xspace}
\newcommand{\veva}{\textsf{Vevacious}\xspace}
\newcommand{\spheno}{\textsf{SPheno}\xspace}

\newcommand{\ddm}{\textsf{DirectDM}\xspace}

\newcommand\xx{\raisebox{0.2ex}{\smaller ++}\xspace}
\newcommand\Cpp{\textsf{C\xx}\xspace}

\newcommand\Python{\textsf{Python}\xspace}
\newcommand\python{\Python}

\newcommand\Fortran{\textsf{Fortran}\xspace}
\newcommand\YAML{\textsf{YAML}\xspace}

\newcommand\cmake{\textsf{CMake}\xspace}

\newcommand\beq{\begin{equation}}
\newcommand\eeq{\end{equation}}

\newcommand{\subparagraph}{} 
\journalname{Eur.\ Phys.\ J.\ C}
\smartqed

\makeatletter
\patchcmd{\ttlh@hang}{\parindent\z@}{\parindent\z@\leavevmode}{}{}
\patchcmd{\ttlh@hang}{\noindent}{}{}{}
\makeatother

\usepackage{pifont}
\newcommand{\cmark}{\ding{51}}%
\newcommand{\xmark}{\ding{55}}%

\usepackage{enumitem}

\usepackage{braket}

\newcommand{\nm}{\metavar{new\_model}}
\newcommand{\pn}{\metavar{particle\_name}}

\begin{document}

\preprintnumber{TTK-21-24, gambit-code-21}

\title{The GAMBIT Universal Model Machine: from Lagrangians to Likelihoods}

\author{Sanjay Bloor\thanksref{imperial,uq,e1} \and
Tom\'as~E.~Gonzalo\thanksref{aachen,monash,e2} \and
Pat Scott\thanksref{imperial,uq,e3} \and
Christopher Chang\thanksref{uq} \and
Are Raklev\thanksref{oslo} \and
Jos{\'e} Eliel Camargo-Molina\thanksref{imperial, uppsala} \and
Anders Kvellestad\thanksref{imperial,oslo} \and
Janina J. Renk\thanksref{imperial,uq,okc} \and
Peter Athron\thanksref{nanjing, monash} \and
Csaba Bal{\'a}zs\thanksref{monash}
}

\institute{
  \gi{imperial}
  \gi{uq}
  \gi{aachen}
  \gi{monash}
  \gi{oslo}
  \gi{uppsala}
  \gi{okc}
  \last{nanjing}
}

\thankstext{e1}{sanjay.bloor12@imperial.ac.uk}
\thankstext{e2}{gonzalo@physik.rwth-aachen.de}
\thankstext{e3}{pat.scott@uq.edu.au\\}

\titlerunning{GUM: The GAMBIT Universal Model Machine}
\authorrunning{Bloor et al. (2020)}

\date{Received: date / Accepted: date}

\maketitle

\begin{abstract}
We introduce the \GB Universal Model Machine (\gum), a tool for automatically generating code for the global fitting software framework \GB, based on Lagrangian-level inputs.  \gum accepts models written symbolically in \fr and \sarah formats, and can use either tool along with \MG and \CH to generate \GB model, collider, dark matter, decay and spectrum code, as well as \GB interfaces to corresponding versions of \spheno, \mo, \pythia and \veva (C\xx).  In this paper we describe the features, methods, usage, pathways, assumptions and current limitations of \gum.  We also give a fully worked example, consisting of the addition of a Majorana fermion simplified dark matter model with a scalar mediator to \GB via \gum, and carry out a corresponding fit.

\end{abstract}

\tableofcontents

\section{Introduction} \label{sec:intro}

The Standard Model (SM) has been exceptionally successful in explaining the fundamental particle physics that underpins the workings of the Universe. Despite these successes, there are still both experimental and theoretical considerations in tension with the SM: the nature of dark matter (DM), the generation of neutrino masses, the hierarchy problem, cosmological matter-antimatter asymmetry, various experimental anomalies (such as flavour or $g-2$), and the stability of the electroweak vacuum. Clearly, the search for physics beyond the SM (BSM) is a multidisciplinary endeavour.

Performing global fits of BSM theories ensures complementarity between experimental and theoretical efforts. This entails consistently predicting multiple observables, comparing them rigorously with experimental results, and using sophisticated statistical sampling techniques in order to obtain meaningful, quantitative inference on theories and their parameters.

The \GB software framework \cite{gambit,grev} aims to address this need, providing a flexible, efficient and scalable code for BSM phenomenology. \GB provides dedicated modules for statistical sampling~\cite{ScannerBit}, DM~\cite{DarkBit}, collider \cite{ColliderBit} flavour \cite{FlavBit} and neutrino \cite{RHN} physics, as well as spectrum generation, decay and precision physics~\cite{SDPBit} and cosmology~\cite{CosmoBit}.

The first version of \GB shipped with various parameterisations of the minimal supersymmetric Standard Model (MSSM) at the weak \cite{MSSM} and grand unified theory (GUT) scales \cite{CMSSM}, a scalar singlet DM extension of the SM~\cite{SSDM} and an effective field theory of flavour interactions \cite{FlavBit,Bhom:2020lmk}. Since then, studies of vacuum stability for scalar singlet DM~\cite{SSDM2}, generic Higgs portal DM models~\cite{HP}, DM effective field theory models~\cite{DMEFT, DMEFT_proceedings}, axions and axion-like particles~\cite{Axions}, additional MSSM models \cite{EWMSSM}, both left- \cite{CosmoBit_numass} and right-handed neutrinos~\cite{RHN}, and inflationary and other cosmological models \cite{CosmoBit} have expanded the \GB model database substantially.

One of the most notable features of \GB is that it is open source. This allows users to study new models and add their own experimental and theoretical constraints to \GB in a modular manner. Although adding a new model to \GB is largely formulaic, it is still not a completely trivial task, as the user still requires some level of understanding of the underlying software design. To this end, we present the \GB Universal Model Machine (\gum): a tool for interfacing symbolic Lagrangian-level tools to \GB, to further automate the procedure of comparing theory to experiment~\cite{Gonzalo:2021cnq}.

Not only does automation increase efficiency and effectiveness in BSM physics, it also reduces the scope for human error, which is inevitably introduced when coding complicated expressions by hand. Development of Lagrangian-level tools has been a very important step in the development of automation in BSM physics.  The original motivation for creating Lagrangian-level tools was to automatically write outputs that could be used for generating matrix element functions, which could in turn be used by Monte Carlo event generators to simulate new physics at particle colliders. The first tool to achieve this was \textsf{LanHEP}~\cite{Semenov:1996es,Semenov:1998eb,Semenov:2002jw,Semenov:2008jy}, originally created to compute vertices for \textsf{CompHEP}~\cite{Boos:1994xb,Pukhov:1999gg,Boos:2004kh} from a simple Lagrangian input. With the release of \fr~\cite{Christensen:2008py,Christensen:2009jx,Christensen:2010wz,Alloul:2013bka}, this quickly expanded to generating output for other matrix element codes, such as \textsf{MadGraph/MadEvent}~\cite{Stelzer:1994ta,Maltoni:2002qb,Alwall:2007st,Alwall:2011uj,Alwall:2014hca}, \CH~\cite{Pukhov:2004ca,Belyaev:2012qa}, \textsf{FeynArts}~\cite{Hahn:1998yk,Hahn:2000kx,Hahn:2000jm,Hahn:2001rv}, \textsf{SHERPA}~\cite{Gleisberg:2008ta} and \textsf{WHIZARD/O'Mega}~\cite{Kilian:2007gr,Moretti:2001zz}. \sarah~\cite{Staub:2008uz,Staub:2009bi,Staub:2010jh,Staub:2012pb,Staub:2013tta,Staub:2015kfa} was also developed around the same time, initially with a particular focus on supersymmetry, but soon expanding to a much larger range of models.

The success of \fr and \sarah in generating Feynman rules for use by matrix element generators lead to the creation of a new filetype, the `Universal FeynRules Output' (\ufo) \cite{Degrande:2011ua}. These \ufo files encode information about the particles, the parameters and interaction vertices for a given model. They can be generated by both \fr and \sarah, and handled by a range of matrix element generators such as \MG, \textsf{GoSam}~\cite{Cullen:2011ac,Cullen:2014yla} and \textsf{Herwig++}~\cite{Bahr:2008pv,Bellm:2015jjp}.

As the search for new physics spans more than just collider physics, it has been necessary for Lagrangian-level tools to generate output for tools in other areas of physics, outside of collider phenomenology.  The \ufo-compatible package \mdm \cite{Backovic:2013dpa,Backovic:2015cra,Ambrogi:2018jqj} has been built on top of \MG, for computing DM relic densities and direct and indirect detection signals. From \sarah, inputs can now also be generated for DM phenomenology with \mo~\cite{Belanger:2001fz,Belanger:2004yn,Belanger:2006is,Belanger:2008sj,Belanger:2010gh,Belanger:2013oya,micromegas}, spectrum generation with \spheno~\cite{Porod:2003um,Porod:2011nf} and \FS~\cite{Athron:2014yba,Athron:2017fvs,Athron:2021kve}, flavour physics observables with \spheno and \textsf{FlavorKit}~\cite{Porod:2014xia}, and calculations of the stability of electroweak symmetry breaking (EWSB) vacuum with \veva~\cite{Camargo-Molina:2013qva}.

Although \fr and \sarah were both created to solve essentially the same problem, they serve different purposes. \fr is concerned with computing Feynman rules for \textit{any} given Lagrangian, including effective ones, and performing physics at tree level. \sarah on the other hand places far more emphasis on renormalisable theories. As a result, any UV-complete model can be implemented in both \fr and \sarah, and any output generated by \fr for such models can also be created by \sarah.  However, \sarah is also able to compute renormalisation group equations (RGEs) at 2-loop order and particle self-energies at 1-loop order, allowing its `downstream beneficiaries' \spheno and \FS to generate corrected mass spectra at the 1-loop level.

\begin{table*}[t!]
  \centering
  \begin{tabular}{l l l l}
  \toprule
  Generated \GB backends                 &  \fr    & \sarah & Usage in \GB \\
  \midrule
  \CH                              & \cmark  & \cmark & Decays, cross-sections \\
  \mo (via \CH)                    & \cmark  & \cmark & DM observables \\
  \pythia (via \MG)                & \cmark  & \cmark & Collider physics \\
  \spheno                          & \xmark  & \cmark & Particle mass spectra, decay widths \\
  \veva C\xx                           & \xmark  & \cmark & Vacuum stability \\
  \bottomrule
  \end{tabular}
  \caption{\GB backends with \gum support and Lagrangian-level tools used to generate them.  Apart from the external packages listed, \gum also produces \GB Core and physics module code tailored to the model and observables of interest.
  }
  \label{tab::outputs}
\end{table*}

Although the outputs of \sarah are more sophisticated than those of \fr, it also has limitations. Unlike in \fr, it is not generally possible to define non-renormalisable theories or higher-dimensional effective theories in \sarah. We therefore provide interfaces to both \fr and \sarah to allow the user to incorporate a vast range of theories into \GB, from effective field theories (EFTs) via \fr to complex UV-complete theories in \sarah. We stress that if a model \emph{can} be implemented in \sarah, then the user \emph{should} use \sarah over \fr\ -- both to use \GB to its full potential, and to perform more detailed physics studies.  The basic outputs available from \gum in each case are summarised in Table~\ref{tab::outputs}.\footnote{Some readers will note the absence of \FS from this list; this is due to the complex \Cpp templates used in \FS and the fact that supporting it fully as a backend in \gambit requires significant development of the classloading abilities of the backend-on-a-stick script (\BOSS) \cite{gambit}. Once this challenge has been overcome, future versions of \gum will also generate code for \flexiblesusy and its other flexi-bretheren. }

This manual is organised as follows: in Sec.~\ref{sec:code}, we describe the code structure and outputs of \gum. In Sec.~\ref{sec:usage} we give usage details, including installation, the \gum file, and particulars of \fr and \sarah model files. In Sec.~\ref{sec:example} we provide a worked example, where we use \gum to add a simplified DM model to \GB, and perform a quick statistical fit to DM observables. Finally, in Sec.~\ref{sec:summary}, we discuss future extensions of \gum and summarise. We include details of the new \GB interfaces to \CH, \veva and \sarah-\spheno (the auto-generated version of \spheno created using \sarah) in the Appendix.

\gum is open source and part of the \gambit \textsf{2.0} release, available from \href{http://gambit.hepforge.org}{gambit.hepforge.org} under the terms of the standard 3-clause BSD license.\footnote{\href{http://opensource.org/licenses/BSD-3-Clause}{http://opensource.org/licenses/BSD-3-Clause}.}

\section{Code design}\label{sec:code}

\GB consists of a set of Core software components, a sampling module \scannerbit \cite{ScannerBit}, and a series of physics modules \cite{ColliderBit,DarkBit,SDPBit,FlavBit,RHN,CosmoBit}.  Each physics module is in charge of a domain-specific subset of \GB's physical calculations. \gum generates various snippets of code that it then adds to parts of the \GB Core, as well as to some of the physics modules, enabling \gambit to employ the capabilities of those modules with the new model.

Within the Core, \gum adds code for any new particles to the \GB particle database, and code for the new model to the \GB models database, informing \GB of the parameters of the new model so that they can be varied in a fit.  \gum also generates interfaces (frontends) to the external codes (backends) that it is able to generate. The backends supported by \gum in this manner are those listed as outputs in Table \ref{tab::outputs}.

Within the physics modules, \gum writes new code for the \specbit \cite{SDPBit} module, responsible for spectrum generation within \GB, \decaybit \cite{SDPBit}, responsible for calculating the decays of particles, \darkbit \cite{DarkBit}, responsible for DM observables, and \colliderbit \cite{ColliderBit}, the module that simulates hard-scattering, hadronisation and showering of particles at colliders, and implements subsequent LHC analyses.

\gum is primarily written in \python, with the exception of the \Mathematica interface, which is written in \Cpp and accessed via \textsf{Boost.Python}.

Initially, \gum parses a \dgum input file, using the contents to construct a singleton \py{gum} object.  Details of the input format can be found in Sec.~\ref{sec:gumfile}.  \gum then performs some simple sanity and consistency checks, such as ensuring that if the user requests DM observables, they have also specified a DM candidate. \gum then opens an interface to either \fr or \sarah via the Wolfram Symbolic Transfer Protocol (\textsf{WSTP}), loads the \fr or \sarah model file that the user has requested into the \Mathematica kernel, and performs some additional sanity checks using the inbuilt diagnostics of each package.

Once \gum is satisfied with the \fr or \sarah model file, it extracts all physical particles, masses and parameters (e.g.\ mixings and couplings). The minimal information required to define a new particle is its mass, spin, color representation, PDG code, and electric charge (if non-self conjugate). For a parameter to be extracted, it must have an associated LHA block in the \fr or \sarah model file, and an index within that block. Additionally for \fr files, the interaction order used in \ufo files must be set.  For details on the syntax required for all required particle and parameter definitions, see Sec.\ \ref{sec:fr_params} for \fr  model files, and Sec.\ \ref{sec:sarah_params} for \sarah model files.

After extracting particle and parameter information, \gum cross-checks that all particles in the new model exist in the \GB particle database, and adds entries if they do not. \gum uses this same particle and parameter information to also write new entries in both the \GB model database and the \specbit module.  All other calculations rely on a combination of new code within \GB and backends.  In the following sections we provide details of the new code generated by \gum in the \GB model database (Sec\ \ref{sec:models}), within \GB physics modules (Sec\ \ref{sec:modules}), and in the form of new backends and their corresponding frontend interfaces in \GB (Sec\ \ref{sec:backends}).

Many of the \GB code outputs are only generated if the user elects to generate relevant new backend codes with \gum.  Details of which backends must be generated with \gum for it to generate different \GB source files can be found in Table~\ref{tab::what_gum_writes}.

At the end of its run, \gum outputs to screen a set of suggested commands for reconfiguring and rebuilding \gambit and the new backends.  It also emits an example \GB input file for running a scan of the new model (\term{yaml_files/}\nm\term{_example.yaml}, where \nm\ is the name of the new model).

\begin{table*}[tpb]
  \centering
  \begin{tabular}{p{1.8cm} p{9.5cm} p{4.6cm}}
    \toprule
    \GB component     & Entries and/or Amendments                                                    &  Required \dgum entry \newline or backend(s)\\
    \midrule
    Models            & \term{src/SpectrumContents/}\nm\term{.cpp}                            &  \\
                      & \term{include/gambit/Models/models/}\nm\term{.hpp}                    &  \\
                      & \term{include/gambit/Models/SimpleSpectra/}
                      \newline\hphantom{,} \nm\term{\_SimpleSpec.hpp} &  \\
    \midrule
    \specbit          & \term{src/SpecBit\_}\nm\term{.cpp}                                    &  \\
                      & \term{include/gambit/SpecBit/SpecBit\_}\nm\term{\_rollcall.hpp}       &  \\
                      & \term{src/SpecBit\_VS.cpp}                                            & \veva and \spheno \\
                      & \term{include/gambit/SpecBit/SpecBit\_VS\_rollcall.hpp}               & \veva and \spheno  \\
    \midrule
    \darkbit          & \term{src/}\nm\term{.cpp}: Dark Matter ID                             & \guminline{wimp_candidate} \\
                      & \term{src/}\nm\term{.cpp}: Process Catalogue                            & \guminline{wimp_candidate} and \CH  \\
                      & \term{include/gambit/DarkBit/DarkBit\_rollcall.hpp}: Direct detection & \guminline{wimp_candidate} and \micromegas \\
    \midrule
    \decaybit          & \term{src/DecayBit.cpp}                                              & \CH or \spheno \\
                      & \term{include/gambit/DecayBit/DecayBit\_rollcall.cpp}                & \CH or \spheno \\
    \midrule
    \colliderbit      & \term{src/models/}\nm\term{.cpp}                                     & 
    \pythia \\
                      & \term{include/gambit/ColliderBit/models/}\nm\term{.hpp}              & 
    \pythia \\
                      & \term{src/ColliderBit\_Higgs.cpp}                                    & \spheno  \\
                      & \term{include/gambit/ColliderBit/ColliderBit\_Higgs\_rollcall.hpp}   & \spheno  \\
    \midrule
    Backends          & \term{src/frontends/CalcHEP\_3\_6\_27.cpp}                           & \CH \\
                      & \term{src/frontends/MicrOmegas\_}\nm\term{\_3\_6\_9\_2.cpp}          & \micromegas \\
                      & \term{src/frontends/SARAHSPheno\_}\nm\term{\_4\_0\_3.cpp}            & \spheno \\
    \bottomrule
  \end{tabular}
  \caption{Details of the new files that \gum writes or modifies in each part of \GB, for the \protect\dgum file entry \protect\guminline{model:}\protect\nm.  Some files are only written or edited when a model-specific version of a particular backend is requested in the \protect\dgum file. This is achieved by including the corresponding option \protect\metavar{backend\_name}\protect\guminline{:true}, e.g.\ \protect\guminline{spheno:true}, \protect\guminline{pythia:true}, etc.  Entries in the rightmost column indicate where this is the case, i.e.\ for rows with an entry present in the rightmost column, the file listed in that row will only be modified/generated if the backend(s) named in the rightmost column are requested in the \protect\dgum file. Where no such entries exist in the right column, the addition or modification of the \GB source is always performed, regardless of the contents of the \protect\dgum file. The entry ``\protect\guminline{wimp_candidate}'' indicates that the output is only generated if the option of the same name is set in the \protect\dgum file, as \protect\guminline{wimp_candidate:}\protect\metavar{pdg}. Note that the entry ``\protect\guminline{decaying_dm_candidate:}\metavar{pdg}'' can be used in place of \protect\guminline{wimp_candidate} if the DM candidate decays instead of annihilates. All filenames containing \protect\nm\ are newly created by \gum; all others are existing files that \gum amends.
  }
  \label{tab::what_gum_writes}
\end{table*}

\subsection{The \GB model database}\label{sec:models}

For every new model requested, \gum adds a new entry to \GB's hierarchical model database. \gum operates under the condition that no model of the same name exists already in the hierarchy, and there are no entries for it in either \textsf{Models}, \specbit or \darkbit, and will throw an error if it does. If the requested model is a new model, \gum creates a new model file \nm\term{.hpp} in the Models directory (see Table~\ref{tab::what_gum_writes}), with the parameters extracted from \fr or \sarah.

In addition to the model file, \gum creates a list of expected contents for the model's particle spectrum object in \term{SpectrumContents/}\nm\term{.cpp}.  This includes not just pole masses of BSM particles, but also the parameters of the model itself, mixing matrices and various SM parameters. \gum also writes a corresponding simple container for the spectrum \nm\term{\_SimpleSpec} that defines functions for accessing the spectrum contents and exposes them to the \gambit~\cpp{Spectrum} class.

\subsection{Modules}\label{sec:modules}

\subsubsection{\specbit}

\specbit includes structures for storing and passing around the so-called mass spectrum, i.e.\ the pole masses, mixings and Lagrangian parameters of the model. If \spheno is used to obtain the spectrum then the $R_\xi$-gauge will be set to $\xi=1$ and the Lagrangian parameters given in the $\overline{MS}$ scheme for non-superymmetric models and the $\overline{DR}^\prime$ scheme for supersymmetric models.  The scale at which these parameters are given will depend on the spectrum generator and will also be stored in the spectrum stored in \specbit.

Following the structure of the simple spectrum container, \gum writes module functions in \specbit that allow the construction of an object of the \cpp{Spectrum} class. The capability \nm\cpp{_spectrum} and its module functions are declared in the header file \term{SpecBit\_}\nm\term{\_rollcall.hpp} and defined in \term{SpecBit\_}\nm\term{.cpp}. The spectrum is either defined directly in terms of phenomenological model parameters, or generated from the Lagrangian parameters using \spheno.

By default, in the absence of a spectrum generator, \gum writes a simple module function in \specbit, \cpp{get_}\nm\cpp{_spectrum}, that fills a \cpp{Spectrum} object with SM values and the input parameters of the model. If using \sarah to generate \GB code, the pole masses of BSM particles are computed using the tree-level relations provided by \sarah. However, these tree-level masses from \sarah are only used for very simple models, such as those without additional Higgs states, as more complicated models include non-trivial mixings and modify the electroweak symmetry breaking (EWSB) conditions. In the latter case a spectrum generator (e.g. \spheno) should be used. When producing \GB output from \fr, however, there are no such relations available, and thus the particle masses are model parameters and the \cpp{Spectrum} object is filled with those.

If the \spheno output is requested from \sarah for a model, \gum writes a module function, \cpp{get_}\nm\cpp{_spectrum_SPheno}, with the backend requirements necessary to generate the full spectrum, with all particle masses, mixing matrices, etc. Hence, for improved precision spectra, it is recommended that the user implement their model using \sarah, and request the spectrum to be provided by \spheno.

If \veva output is requested, for each new BSM model \gum writes new model-specific code in the \specbit vacuum stability file, \term{SpecBit/src/SpecBit\_VS.cpp}, and adds appropriate entries to the corresponding rollcall header. \gum provides two new module functions to interact with \veva. Firstly, \cpp{prepare_pass_}\nm\cpp{_spectrum_to_vevacious} with capability \cpp{pass_spectrum_to_vevacious}, which interfaces the \cpp{Spectrum} object to the \veva object. Secondly, \cpp{vevacious_file_location_}\nm, which directs \GB to the location of the input \veva files generated by \sarah.

\subsubsection{\decaybit}

Whenever decay information is requested for a new model, \gum amends the header \term{DecayBit\_rollcall.hpp} and source \term{DecayBit.cpp} files to add the decays of the particles in the model. The information for the decays can be provided separately by two backends in the \gum pipeline: \CH and \spheno.

\CH generates tree-level decays for each new BSM particle, plus new contributions to any existing particles in \decaybit such as the SM Higgs and the top quark. \gum adds these to \decaybit by adding the new decay channels wherever possible to any existing \cpp{DecayTable::Entry} provided by a module function with capability \pn\cpp{_decay_rates}.  If no such function exists, it instead creates a new module function \cpp{CH_}\nm\cpp{_}\pn\cpp{_decays}, with this capability-type signature, where \pn\ comes from the \sarah/\fr model file. \gum then modifies the module function \cpp{all_decays} to add the decays of any particles for which it has written new module functions.  \gum presently only supports the generation of $1\to 2$ decays, as support for 3-body decays within \decaybit is currently limited to fully integrated partial widths; consistent inclusion of integrated partial widths provided by \spheno and differential decay widths provided by \CH into \decaybit will require further development of the \cpp{DecayTable} infrastructure, and a careful treatment of integration limits. Note also that \gum does not currently write any BSM contribution for $W$ and $Z$ boson decay via the \CH interface, but this is planned for future releases.

\spheno by default computes leading order decays (tree-level or one-loop) for all BSM and SM particles, and adds universal higher-order corrections for specific decays, which are very important for Higgs decays. In addition it provides an alternative computation of full one-loop decays for all particles. The choice of method is left to the user via the \spheno option \yaml{OneLoopDecays}. As \spheno provides decay widths for all particles in the spectrum, \gum creates a new module function \cpp{all_}\nm\cpp{_decays_from_SPheno}, which returns a \cpp{DecayTable} filled with all decay information computed by \spheno.

The default behaviour of \gum is to ensure that it always generates decay code of some sort when needed. This ensures that a complete \cpp{DecayTable} for the new model can be provided within \GB for dependency resolution, by providing the capability \cpp{all_decays} in \decaybit. A viable \GB\ \cpp{DecayTable} is required for the functioning of many external codes such as \pythia and \mo: if any new particle is a mediator in a process of interest, then its width is needed.  To this end, \gum activates \CH output automatically if decays are needed by other outputs that have been activated in the \dgum file, but neither \CH nor \spheno output has been explicitly requested.

\subsubsection{\darkbit}

If the user specifies a Weakly-Interacting Massive Particle (WIMP) DM candidate for a model\footnote{We note that it is not currently possible to declare more than one DM candidate, and \gum will fail if attempted. Additionally, we do not recommend specifying a single DM candidate for models with multi-component DM, as this could lead to inconsistent results.}, \gum writes the relevant code in \darkbit. Each individual model tree in \darkbit has its own source file, so \gum generates a new source file \term{src/}\nm\term{.cpp}, and amends the \darkbit rollcall header accordingly. At a minimum, \gum includes a new module function \cpp{DarkMatter_ID_}\nm\ in this file. It then adds the remainder of the source code according to which backends the user selects to output code for in their \dgum file; currently available options are \CH and \mo.

If the user requests \CH output, then a new module \cpp{TH_ProcessCatalog_}\nm\ providing the Process Catalogue is written.  The Process Catalogue houses all information about annihilation and decay of DM, and decays of all other particles in the spectrum.  All computations of indirect detection and relic density likelihoods in \darkbit begin with the Process Catalogue. For details, see Sec. 6.3 of~\cite{DarkBit}. All processes that \gum adds to the Process Catalogue are $2\to2$ processes computed at tree level by \CH.

The Process Catalogue is used to compute the relic abundance of annihilating DM via the \darkbit interface to the Boltzmann solver in \ds.  The Process Catalogue interface does not currently fully support co-annihilations nor 3-body final states, so \gum does not generate these processes. Such functionality is planned for future releases of \darkbit, and will be supported by \gum at that time.  Note however that \gum \textit{does} support the full generation of \mo versions including these effects via the \CH-\mo interface. If co-annihilations or three-body final states are expected to be important in a new physics model, the user should therefore use \mo to compute the relic abundance from \darkbit, in preference to the \ds Boltzmann solver.

For decaying DM candidates, the user would need to implement their own relic density calculation, as appropriate for the specific model in question.  Although \darkbit can calculate spectral yields from DM decay, at present the likelihoods for indirect detection in \darkbit do not support decaying dark matter (in large part because neither \gamlike nor \nulike presently support decaying DM).  Existing direct detection likelihoods can still be used out of the box without any relic density calculation, if the user assumes that the decaying DM candidate constitutes all of the DM.

When writing \mo output, \gum adds new entries to the \cpp{ALLOW_MODELS} macro for existing \mo functions in \darkbit. To use \mo' relic density calculator, \gum adds an entry to the module function \cpp{RD_oh2_Xf_MicrOmegas}. For more information, see Sec.\ 4.2 of Ref.\ \cite{DarkBit}.  \gum also provides an interface to the module function \cpp{DD_couplings_MicrOmegas}, which returns a \cpp{DM_nucleon_couplings} object containing the basic effective spin-independent and spin-dependent couplings to neutrons and protons $G_{\rm SI}^{\rm p}$, $G_{\rm SI}^{\rm n}$, $G_{\rm SD}^{\rm n}$, and $G_{\rm SD}^{\rm n}$. This object is readily fed to \ddcalc for computing likelihoods from direct detection experiments. For more information, see Sec.\ 5 of Ref.\ \cite{DarkBit}.

For more complicated models where the standard spin-independent and spin-dependent cross-sections are not sufficient, \mo is not able to compute relevant couplings. In this case, the user should perform a more rigorous calculation of WIMP-nucleon (or WIMP-nucleus) couplings by alternative means. This is required when, for example, scattering cross-sections rely on the momentum exchange between incoming DM and SM particles, or their relative velocity.  The full set of 18 non-relativistic EFT (NREFT) DM-nucleon operators are defined in both \darkbit and \ddcalc \textsf{2}, and described in full in the Appendix of Ref.~\cite{HP}. These operators fully take into account velocity and momentum transfer up to second order, and should typically be used in cases where the entirety of the physics is not captured by just $\sigma_{\rm{SI}}$ and $\sigma_{\rm{SD}}$.  Whilst the new 2.1 version of \gambit \cite{DMEFT} allows for automated translation of high-scale relativistic effective DM-parton couplings to low-scale NREFT couplings via an interface to \ddm \cite{Bishara:2017nnn,Brod:2017bsw}, there is no established automated matching procedure for connecting other high-scale models (as defined in \fr or \sarah model files) to the Wilson coefficients of the relativistic EFT. \gum therefore does not automatically write any module functions connecting the \GB\ \cpp{Spectrum} object to the NREFT interface of \ddcalc; once such a procedure exists, \gum will be extended accordingly.

\subsubsection{\colliderbit} \label{sec:colliderbit}

In \colliderbit, simulations of hard-scattering collisions of particles are performed using the Monte Carlo event generator \pythia~\cite{Sjostrand:2014zea}. These events are passed through detector simulation and then fed into the \GB analysis pipeline, which predicts the signal yields for the new model.  These can then be compared to the results of experimental searches for new particles.

For a new BSM model, the matrix elements for new processes unique to the model must be inserted into \pythia in order for it to be able to draw Monte Carlo events from the differential cross-sections of the model. To achieve this, \gum communicates with \madgraph to generate matrix element code for \pythia, and writes the appropriate patch to insert it into \pythia. Alongside the matrix elements, this \pythia patch also inserts any newly defined LHA blocks.

When \pythia output is requested, \gum writes a series of new \colliderbit module functions in the source file \term{ColliderBit/src/models/}\nm\term{.cpp}, and a corresponding rollcall header file.  The new functions give \colliderbit the ability to \begin{itemize}
\item[i)] collect the relevant \cpp{Spectrum} and \cpp{DecayTable} objects from other modules and provide them to the newly-generated copy of \pythia (capability \cpp{SpectrumAndDecaysForPythia}, function \cpp{getSpectrumAndDecaysForPythia_}\nm),
\item[ii)] initialise the new \pythia for Monte Carlo events (capability \cpp{HardScatteringSim}, function \cpp{getPythia_}\nm), and
\item[iii)] call the new \pythia in order to generate a Monte Carlo event (capability \cpp{HardScatteringEvent}, function \cpp{generateEventPythia_}\nm).
\end{itemize}

In addition to the likelihood from LHC new particle searches, \colliderbit also provides likelihoods associated with Higgs physics. This is done via interfaces to \higgsbounds~\cite{Bechtle:2008jh,Bechtle:2011sb,Bechtle:2013wla,arXiv:1507.06706} and \higgssignals~\cite{HiggsSignals,Bechtle:2014ewa}, which use information on Higgs signal rates and masses from the Tevatron, LEP and the LHC. When a new model is added to \GB with \gum, if \spheno output is requested from \sarah, \gum constructs a new \cpp{HiggsCouplingsTable} used as input to the Higgs likelihoods, and amends the appropriate module function entries in \term{ColliderBit\_Higgs\_rollcall.hpp} and \term{ColliderBit\_Higgs.cpp}.

\subsection{Backends}\label{sec:backends}

In the \gambit framework, backends are external tools that \gambit links to dynamically at runtime, in order to compute various physical observables and likelihoods. Out of the full list of backends that can be interfaced with \gambit, a small selection of them can work for generic BSM models. In particular, \gum is able to produce output for \spheno~\cite{Porod:2003um,Porod:2011nf}, \veva~\cite{Camargo-Molina:2013qva}, \CH~\cite{Belyaev:2012qa}, \micromegas~\cite{micromegas}, \pythia~\cite{Sjostrand:2014zea}, \higgsbounds~\cite{Bechtle:2008jh,Bechtle:2011sb,Bechtle:2013wla,arXiv:1507.06706} and \higgssignals~\cite{HiggsSignals,Bechtle:2014ewa}. Thus, we briefly describe here the specific outputs generated by \gum for each of these backends, along with and any corresponding \gum and \GB \YAML input file entries needed to use them. Unless otherwise stated, \gum has been developed to work with specific versions of the backends.

\subsubsection{(\sarah-)\spheno \textsf{4.0.3}}

\textbf{Required \dgum file entry:} \guminline{spheno: true}\\

\spheno is a spectrum generator capable of computing one-loop masses and tree-level decay branching fractions in a variety of BSM models. The model-specific code is generated by \sarah and combined with the out-of-the-box \spheno code into a single backend for \gambit. For each model \gum thus provides an interface between \gambit and the \spheno version via a new frontend, \term{SARAHSPheno\_}\nm\term{\_4\_0\_3.cpp}. Details about this interface, which differs significantly from the \spheno interface described in \cite{SDPBit}, can be found in Appendix \ref{app:spheno}.

In order to generate \spheno output from \sarah, the user must provide a \term{SPheno.m} file in the same directory as the \sarah model files. For details of the contents of these files, we refer the reader to the \sarah manual~\cite{Staub:2015kfa}.

Once the appropriate \gambit code is generated by \gum, the new capability \nm\cpp{_spectrum} is added to \specbit to compute the spectrum using \spheno. The new \spheno generated \cpp{Spectrum} object can be obtained for a specific model in a run via the \YAML entry in the \gambit input file:
\begin{lstyaml}
ObsLikes:
  - purpose: Observable
    capability: @\nm@_spectrum
\end{lstyaml}
As usual, if more than one module function can provide the same capability, as can happen, for example, if \FS is also present, the \spheno specific one can be selected by the rule in the \gambit input file.

\begin{lstyaml}
Rules:
  - capability: @\nm@_spectrum
    function:   get_@\nm@_spectrum_SPheno
\end{lstyaml}

In addition to their masses and mixings, \spheno can compute the tree-level decay branching fractions for all particles in the spectrum, including some radiative corrections to the decays of Higgs bosons. The \gum-generated code in \decaybit includes the new module function \cpp{all_}\nm\cpp{_decays_from_SPheno}, which returns a \gambit~\cpp{DecayTable} as computed by \spheno. This provides an alternative to the usual \cpp{all_decays} function in \decaybit.  When run with default settings, \gambit will preferentially select \cpp{all_}\nm\cpp{_decays_from_SPheno} whenever possible, rather than \cpp{all_decays}, as the former is more model-specific.  If necessary, the user can also manually instruct \gambit to use this function by specifying a rule for the \cpp{decay_rates} capability:
\begin{lstyaml}
Rules:
  - capability: decay_rates
    function:   all_@\nm@_decays_from_SPheno
\end{lstyaml}

To build a newly-added \sarah-generated \spheno (``\sarah-\spheno'') within \gambit, the appropriate commands to run in the \gambit build directory are
\begin{lstterm}
cmake ..
make sarah-spheno_@\nm@
\end{lstterm}
where the backend name \term{sarah-spheno} is used to differentiate the corresponding code from the out-of-the-box version of \spheno.

\subsubsection{\veva \textsf{1.0} (C\xx)} \label{sec:veva_desc}

\textbf{Required \dgum entries:} \guminline{vevacious: true}, \guminline{spheno: true}.\\

\veva computes the stability of the scalar potential for generic extended scalar sectors~\cite{Camargo-Molina:2013qva}. A recent C\xx version \cite{vevaciousgithub} has recently been interfaced to \GB as a backend \cite{VS_GUT} and is the one used by \gum. The \gambit interface to \veva is explained in more detail in Appendix~\ref{app:vevacious}.

To test the stability of the EWSB vacuum, Vevacious checks whether other deeper minima exist, in which case it computes the tunnelling probability to either the nearest (to the EWSB vacuum) or the deepest of such minima. The user can select whether to compute the tunnelling probability for either of them by using the \yaml{sub_capabilities} options in the \gambit input entry for the capability \cpp{VS_likelihood} as
\begin{lstyaml}
ObsLikes:
  - purpose:      LogLike
    capability:   VS_likelihood
    sub_capabilities:
      - global
      - nearest
\end{lstyaml}
If both minima are chosen, \veva computes the probability of tunnelling to both if they are different. The capability \cpp{compare_panic_vacua} in \GB checks if the nearest minimum is also the global minimum. If that is the case, the transition to the minimum is only computed once, which reduces the computation time significantly.
In many instances, such as for the MSSM, the tunneling path optimization within \veva can be a very time-consuming step, therefore computing the tunnelling probability to both minima is not always recommended, as it requires running \veva twice for parameter points where the minima are different, and thus can be prohibitively slow for very large scalar sectors.

For each minimum, \veva by default computes both the zero-temperature (\textit{quantum}) tunnelling probability as well as the finite-temperature (\textit{thermal fluctuation}) probability. To select \yaml{quantum} or \yaml{thermal}, it is possible to provide options to the \yaml{sub_capabilities} as
\begin{lstyaml}
ObsLikes:
  - purpose:      LogLike
    capability:   VS_likelihood
    sub_capabilities:
      global: [quantum]
\end{lstyaml}
If both minima are selected, the same tunnelling strategy must be selected for both. \GB computes the likelihood by combining the decay widths for all (independent) transitions as reported by \veva.

For each new model, \sarah generates the model files required by \veva and \gum moves them into the patch directory in \GB. Note that if the user wishes to request \veva output, they must also request \spheno output (and provide a \term{SPheno.m} file). This is due to the \gum interface utilising \Mathematica symbols provided by \sarah's \spheno routines.

Once a new model has been generated by \gum, \veva can be built from within the \gambit build directory with the command
\begin{lstterm}
make vevacious
\end{lstterm}
which will either download and build \veva if it is not installed, or simply move the new model files from the \GB patch directory to the \veva directory if it is already built. Note that building \veva for the first time will also download and install \textsf{MINUIT}~\cite{James:1975dr}, \textsf{PHC}~\cite{Verschelde2011}, and \textsf{HOM4PS2}~\cite{Lee2008}.

\subsubsection{\CH \textsf{3.6.27}}

\textbf{Required \dgum file entry:} \guminline{calchep: true} \\
\textbf{Optional:} \guminline{wimp_candidate:}\metavar{pdg} for annihilating DM, or \hphantom{\textbf{Optional:}} \guminline{decaying_dm_candidate:}\metavar{pdg} for decaying DM \\

\gum uses the backend convenience function \cpp{CH_Decay_Width} provided by the new \CH frontend (described in Appendix~\ref{app:calchep}), to compute tree-level decay widths.

For each new BSM decay, \gum generates a model-specific \cpp{DecayTable::Entry}. For each newly-added decaying particle, \gum writes a module function \cpp{CH_}\nm\cpp{_}\pn\cpp{_decays}, which requires the ability to call the backend convenience function \cpp{CH_Decay_Width}. All new decays are then gathered up by the existing \decaybit function \cpp{all_decays}, which \gum modifies by adding an \cpp{if(ModelInUse(}\nm\cpp{))} switch for newly-added decaying particles in the new model.

The appropriate \gambit input rule for a \CH-generated \cpp{DecayTable} is simply
\begin{lstyaml}
Rules:
  # Use DecayBit (and CalcHEP), for decay
  # rates; not an SLHA file or SPheno
  - capability: decay_rates
    function:   all_decays
\end{lstyaml}

If the user specifies the PDG code of a WIMP candidate via \guminline{wimp_candidate:}\metavar{pdg} or \guminline{decaying_dm_candidate:}\metavar{pdg}, then \gum creates a \darkbit entry for the new model.

In the case of \guminline{wimp_candidate:}\metavar{pdg}, \gum utilises the backend convenience function \cpp{CH_Sigma_V} provided by the \CH frontend, to build the Process Catalogue. It does this by computing $2\rightarrow2$ scattering rates as a function of the relative velocity $v_{\rm{rel}}$, which are in turn fed to the appropriate module functions.

In the case of \guminline{decaying_dm_candidate:}\metavar{pdg}, \gum instead utilises the branching ratios from the \cpp{DecayTable::Entry} for the decaying DM candidate, which can be constructed from \CH, as described above.

The information contained within the Process Catalogue can be used by the \GB native relic density solver for annihilating DM (using the function \cpp{RD_oh2_general} via \darksusy), and for all indirect detection rates.  For annihilating DM, these utilise the velocity weighted annihilation cross-section $\sigma v_\text{rel}$. This is usually evaluated at $v_{\rm{rel}}=0$ (such as in the case of $\gamma$ rays), but for solar neutrinos, $v_{\rm{rel}}$ is set to the solar temperature $T_{\rm{Sun}}$.  For decaying DM, indirect detection rates are explicitly disabled until backend support for decaying DM becomes available.

In the first release of \gum, there is no support for 4-fermion interactions in \CH, as neither \fr nor \sarah is able to produce output for these.

From the \GB build directory the command
\begin{lstterm}
make calchep
\end{lstterm}
will build \CH if it is not installed; otherwise it will move the new \CH model files from the \GB patch directory to the \CH model directory.

\subsubsection{\micromegas \textsf{3.6.9.2}}

\textbf{Required \dgum file entry:} \guminline{micromegas: true}\\
\textbf{Optional:} \guminline{wimp_candidate:}\metavar{pdg} for annihilating DM, or \hphantom{\textbf{Optional:}} \guminline{decaying_dm_candidate:}\metavar{pdg} for decaying DM \\

\textsf{MicrOMEGAs} is a code capable of computing various DM observables for BSM models with WIMP candidates, such as the relic abundance, direct detection cross-sections, and indirect detection observables. Each \mo installation in \GB is a separate backend, as the source is compiled directly with the model files. Therefore for each newly-added \mo model, \gum creates a new single backend for \mo, via the new frontend \term{MicrOmegas\_}\nm\term{\_3\_6\_9\_2.cpp}.

\textsf{MicrOMEGAs} uses \CH files as input so is subject to the same caveats as \CH, covered above. \textsf{MicrOMEGAs} assumes that there is an additional symmetry under which the SM is even and any dark matter candidate is odd. \textsf{MicrOMEGAs} distinguishes an odd particle by having its name begin with a tilde, such as \mathematica{\~chi}. If no particle name in a theory begins with a tilde, the \mo routines will fail.\footnote{If the DM particle is not self-conjugate, its antiparticle should also begin with a tilde.} The particle name is set by the \mathematica{ParameterName} option in \fr, and the \mathematica{OutputName} option in \sarah. If the indicated DM candidate does not have a particle name beginning with a tilde, \gum throws an error.

\gum provides a simple interface to the relic density calculation in \mo, in the case of annihilating DM.  The \gambit input entry for computing the relic density with \mo is:
\begin{lstyaml}
Rules:
  - capability: RD_oh2
    function:   RD_oh2_MicrOmegas
\end{lstyaml}
If the user provides a decaying DM candidate, then \gum does not provide an interface to the \mo relic density routines.  If the user wishes to compute the relic density for a decaying DM candidate, they must implement their own relic density calculation by hand.

\gum also provides a simple interface to the direct detection routines in \mo, which simply provide calculations of the spin-independent and spin-dependent cross-sections, which are added for both annihilating and decaying DM.  This is fed to \ddcalc~\cite{DarkBit,HP} which computes expected rates for a wide range of direct detection experiments.

As each installation of \mo is a separate backend, each requires a specific build command to be run in the \gambit build directory:
\begin{lstterm}
make micromegas_@\nm@
\end{lstterm}

Future versions of \GB and \gum will interface to \mo \textsf{5}~\cite{Belanger:2018ccd}, which contains routines for computing the relic abundance of DM via freeze-in, and allows for two-component DM.

\subsubsection{\pythia \textsf{8.212}} \label{sec:pythia}

\textbf{Required \dgum entries:} \guminline{pythia: true}, \guminline{collider_processes: [...]}.\\

If the user requests \pythia output for a given model, either \fr or \sarah generates a collection of \ufo files. \gum calls \MG directly using the \ufo model files, and generates new output for \pythia in the form of matrix elements. \gum then writes the appropriate entries in the backend patch system and connects the new matrix elements with those existing in \pythia, and adds the corresponding entry to the file \term{default\_bossed\_versions.hpp}.  Because \pythia is a \Cpp code, and the \cpp{Pythia} class it defines is used directly in \colliderbit, new versions of the backend must be processed for automated classloading from the corresponding shared library by \BOSS (the backend-on-a-stick script \cite{gambit}).  This process generates new header files that must be \term{#include}(d) in \GB itself, and therefore picked up by \cmake. Therefore, a new version of \pythia is correctly built by running the commands
\begin{lstterm}
cmake ..
make pythia_@\nm@
cmake ..
make -j@\metavar{n}@ gambit
\end{lstterm}
in the \GB build directory, where \metavar{n} specifies the number of processes to use when building.
In the current version of \colliderbit, functions from \nulike \cite{IC22Methods,IC79_SUSY} are also required in order to perform inline marginalisation over systematic errors in the likelihood calculation; this can be built with
\begin{lstterm}
make -j@\metavar{n}@ nulike
\end{lstterm}
also in the \GB build directory.

The user must provide a list of all processes to include in the new version of \pythia in the \dgum file under the heading \guminline{collider_processes}; see Sec.~\ref{sec:gumfile} for details.

Once a new \pythia has been created, it has access to all implemented LHC searches within \colliderbit. The relevant \gambit input file entry to include \pythia simulations is
\begin{lstyaml}
ObsLikes:
  # LHC likelihood from Pythia
  - purpose:      LogLike
    capability:   LHC_Combined_LogLike
\end{lstyaml}
along with rules specifying how to resolve the corresponding dependencies and backend requirements, e.g.
\begin{lstyaml}
Rules:
  # Choose LHC likelihood form (assume normal or
  # log-normal distribution for systematics)
  - capability: LHC_LogLikes
    backends:
    - {capability: lnlike_marg_poisson_gaussian_error}

  # Choose to get cross-sections by Monte Carlo
  - capability: TotalCrossSection
    function: getEvGenCrossSection_as_base

  # Just use unweighted cross-sections
  - capability: EventWeighterFunction
    function: setEventWeight_unity

  # Select where to import model-specific decays
  # from - CalcHEP in this instance
  # (Alternatively: all_@\nm@_decays_from_SPheno)
  - capability: decay_rates
    function: all_decays
\end{lstyaml}

The matrix elements generated by \MG can include extra hard partons in the final state so that jet-parton matching is needed to avoid double counting between the matrix elements and the parton shower in \pythia. Currently, this is not automatic in \gum and must be implemented by the user as needed.

Traditional MLM matching, e.g.\ as found in \madgraph~\cite{Alwall:2011uj}, applies a $k_T$ jet measure cut on partons (\term{xqcut}) at the matrix element level, and separates events with different multiplicities. The optimal value of this cut should be related to the hard scale of the process, e.g.\ the mass of the produced particles, and tuned to ensure smoothness of differential jet distributions and invariance of the cross-section.

The hard scattering events are then showered and a jet finding algorithm (the $k_T$-algorithm~\cite{Catani:1993hr,Ellis:1993tq} implemented in \pythia's \cpp{SlowJet} class in our case) is used on the final state partons to match the resulting jets to the original partons from the hard scatter.
A jet is considered to be matched to the closest parton if the jet measure $k_{T}(\text{parton},\text{jet})$ is smaller than a cutoff \term{qCut}. This parton shower cut, \term{qCut}, should be set slightly above \term{xqcut}. The event is rejected unless each jet is matched to a parton, except for the highest multiplicity sample, where extra jets are allowed below the $k_{T}$ scale of the softest matrix element parton in the event.

While a full implementation of this matching procedure for use in \gum is still in development, to perform simple jet matching for models generated with \gum, the user can make use of the MLM matching machinery already present in \pythia. This can be accessed by the \colliderbit initialisation of \pythia to allow the relevant jet matching inputs to be passed through \pythia settings in the \GB \YAML file, for example:

\begin{lstyaml}
Rules:
  - capability:  HardScatteringSim
    type: Py8Collider_@\nm@_defaultversion
    function: getPythia_@\nm@
    options:
      LHC_13TeV:
        xsec_veto:  0.028
        pythia_settings:
          # Specify MLM matching method
          - JetMatching:merge = on
          - JetMatching:scheme = 1
          - JetMatching:setMad = off
          - JetMatching:jetAlgorithm = 2
          - JetMatching:slowJetPower = 1

          # Jet finding properties
          - JetMatching:coneRadius = 1.0
          - JetMatching:etaJetMax = 5.0

          # Only light flavours in matching
          - JetMatching:nQmatch = 4

          # Maximum number of jets
          # as defined in the Matrix Elements
          - JetMatching:nJetMax = 1

          # Minimal kT for a PYTHIA jet
          - JetMatching:qCut = 30.0

          # Phase space cut to approximate
          # Matrix Element cuts
          - PhaseSpace:pTHatMin = 20.0
\end{lstyaml}

Using the above approach one can approximate the matching for a single extra hard parton in dark matter pair production, by applying the matching cuts  only  on the \pythia side, on events from the matrix elements generated by \madgraph.
Here, we also apply a $p_T$ cut on the partons in the hard scatter in \pythia (\yamlvalue{pTHatMin}). While this $p_T$ cut and the \term{xqcut} are not equivalent, the difference is small for single-jet events because the geometrical part of the $k_T$ jet measure becomes unimportant.

A side-effect of using \yamlvalue{pTHatMin} is that it applies to all final state particles, and so it must be set well below any $\slashed{E}_T$ cuts in the analysis. Potential poor initial cross-section estimates in \pythia  may lower the phase space selection efficiency, inflating computation time. To combat this, \yamlvalue{pTHatMin} may be raised, requiring a sensible balance between efficiency and analysis cut constraints. Due to the limitations of this approach, the accuracy of jet matching cannot be guaranteed, and should be confirmed on a per-model, per-analysis basis.

We refer the reader to the \colliderbit manual~\cite{ColliderBit} for additional details on the \pythia options used within the \GB \YAML file.

The particle numbering scheme used by both \GB and \pythia is that of the PDG.  For dark matter particles to be correctly recognised as invisible by both libraries, their PDG codes must be within the range $51-60$.  Other particles that \pythia and \GB tag as invisible are the SM neutrinos, neutralinos, sneutrinos, and the gravitino.  Where possible, all particles in \sarah and \fr files passed to \gum by the user should adhere to the PDG numbering scheme.  For more details, see Sec.\ 43 of the PDG review~\cite{PDG18}.

\gum checks that any newly-added particle in the \GB particle database is consistent with the definition in \pythia. If there is an inconsistency between the two, \gum will throw an error.  For example, the PDG code 51 is not filled in the \GB particle database by default, but is reserved for scalar DM in \pythia.  \gum will throw an error if the user attempts to add a new particle with PDG code 51 but with spin $1/2$.

\subsubsection{\higgsbounds \textsf{4.3.1} \& \higgssignals \textsf{1.4.0}}

\textbf{Required \dgum file entry:} \guminline{spheno: true}\\

Another advantage of using \spheno to compute decays is that all relevant couplings for \higgsbounds and \higgssignals are automatically computed. Whenever \spheno output is generated, \gum also generates an interface to the \GB implementations of \higgsbounds and \higgssignals via the \GB type \cpp{HiggsCouplingsTable}.

\gum achieves this by generating a function that produces an instance of the \GB native type \cpp{HiggsCouplingsTable} from the decay output of \spheno. The \cpp{HiggsCouplingsTable} object provides all decays of neutral and charged Higgses, SM-normalised effective couplings to SM final states, branching ratios to invisible final states, and top decays into light Higgses. For more details, we refer the reader to the \specbit manual~\cite{SDPBit}.

\GB categorises models into two types: `SM-like' refers to models with only the SM Higgs plus other particles, and `MSSM-like' refers to models with extended Higgs sectors. The appropriate type is automatically selected for each model by the \GB dependency resolver, by activating the relevant one of the module functions in \colliderbit that can provide capability \cpp{HB_ModelParameters}.

For `SM-like' models, \gum edits the \colliderbit module function \cpp{SMLikeHiggs_ModelParameters} to simply pass details of the single Higgs boson from the \cpp{Spectrum} object of the new model. For `MSSM-like' models, \gum edits the \colliderbit function \cpp{MSSMLikeHiggs_ModelParameters}, which communicates the properties of all neutral and charged Higgses to \higgsbounds/\higgssignals in order to deal with extended Higgs sectors.

To ensure the interface to the \cpp{HiggsCouplingsTable} works as expected, the user should make sure that the PDG codes of their Higgs sector mimics those of both \GB and the SLHA:
\begin{lstgum}
# CP-even neutral Higgses
h0: [25, 35, 45]
# CP-odd neutral Higgses
A0: [36, 46]
# Charged Higgs
Hpls: 37
\end{lstgum}

The \cpp{MSSMLikeHiggs_ModelParameters} function automatically supports all MSSM and NMSSM models within \GB, as well as any model with a similar Higgs sector (\eg a Two-Higgs Doublet Model or any subset of the NMSSM Higgs sector). If the user has extended Higgs sectors beyond this, i.e.\ with more Higgses than the NMSSM, then they will need to extend both \gum and \GB manually.

On the \GB side, if the Higgs sector has multiple charged Higgses, more than three CP-even or more than two CP-odd neutral Higgses, the user must write a new function in \term{ColliderBit/src/ColliderBit\_Higgs.cpp} to construct the \cpp{HiggsCouplingsTable} correctly. If there are new CP-even Higgses, this will also require a new entry in \term{Elements/src/smlike_higgs.cpp} to determine the `most SM-like' Higgs.

In \gum, the user must add the PDG codes of additional mass eigenstates to the function \cpp{get_higgses} in \textinline{gum/src/particledb.py} under the appropriate entries \py{neutral_higgses_by_pdg} and \py{charged_higgses_by_pdg}, and also make appropriate changes to the functions \py{write_spectrum_header} in \textinline{gum/src/spectrum.py} to reflect any changes to the construction of the \cpp{HiggsCouplingsTable}.

The appropriate \gambit input entries for using \higgsbounds and \higgssignals likelihoods are simply
\begin{lstyaml}
ObsLikes:
  # HiggsBounds LEP likelihood
  - purpose:      LogLike
    capability:   LEP_Higgs_LogLike

  # HiggsSignals LHC likelihood
  - purpose:      LogLike
    capability:   LHC_Higgs_LogLike
\end{lstyaml}
where the choice of function fulfilling the capability \cpp{HB_ModelParameters} is automatically taken care of by the dependency resolver.

\higgsbounds and \higgssignals can both be built with
\begin{lstterm}
make higgsbounds higgssignals
\end{lstterm}
but neither actually needs to be rebuilt once a new model is added by \gum.

\section{Usage} \label{sec:usage}

\subsection{Installation}\label{sec:build}

\gum is distributed with \GB \textsf{2.0.0} and later. The program can be found within the \term{gum/} folder in the \GB root directory, and makes use of \cmake. In addition to the minimum requirements of \GB itself, \gum also requires at least
\begin{itemize}
 \item \Mathematica \textsf{7.0}
 \item \python \textsf{2.7} or \python 3
 \item The \python \textsf{future} module
 \item Version \textsf{1.41} of the compiled \textsf{Boost} libraries \textsf{Boost.Python}, \textsf{Boost.Filesystem} and \textsf{Boost.System}
 \item \textsf{libuuid}
 \item \textsf{libX11} development libraries.
\end{itemize}

Note that all \cmake flags used in \gum are entirely independent from those used within \GB. From the \GB root directory, the following commands will build \gum:
\begin{lsttermalt}
cd gum
mkdir build
cd build
cmake ..
make -j*@\metavar{n}@*
\end{lsttermalt}
where \metavar{n} specifies the number of processes to use when building. 

\subsection{Running \gum} \label{sec:running_gum}

The input for \gum is a \dgum file, written in the \YAML format. This file contains all of the information required for \gum to write the relevant module functions for \GB, in a similar vein to the input \YAML file used in \GB. \gum is executed with an initialisation file \nm\term{.gum} with the \term{-f} flag, as in
\begin{lstterm}
./gum -f @\nm@.gum
\end{lstterm}

The full set of command-line flags are:
\begin{itemize}
\item[] \term{-d}/\term{--dryrun} \\
      to perform a dry run
\item[] \term{-h}/\textinline{--help} \\
      to display help
\item[] \term{-f}/\term{--file} \metavar{file.gum} \\
      to use the instructions from \metavar{file.gum} to run \gum
\item[] \term{-r}/\term{--reset} \metavar{file.mug} \\
      to use the instructions from \metavar{file.mug} to run \gum in reset mode
\end{itemize}
There are three operational modes of gum: \textit{dry run}, \textit{regular} and \textit{reset}. During a \textit{dry run}, no code is actually written to \GB.  \gum checks that the \Mathematica model file (either \fr or \sarah) is suitable for use, and writes a number of proposed source files for \GB,  but does not actually copy them to the \GB source directories.  This mode can be used for safe testing of new \dgum and model files, without modifying any of \GB.

A \textit{regular} run of \gum will perform all necessary checks, add the new model to \GB and generate all relevant \GB code requested in the \dgum file. After a regular \gum execution, \gum prints a set of commands to standard output for the user to run. It is recommended that the user copies these commands and runs them as instructed, as the order of the suggested build and \cmake steps can be important, due to new templated \textsf{C++} types being provided by backends (currently just \pythia).

In addition to the above, \gum outputs a reset (\term{.mug}) file after a successful run. This file is used in the \textit{reset} mode, and enables the user to remove a \gum-generated model from \GB. Hence, after adding a new model, the user can run the command
\begin{lstterm}
./gum -r @\nm@.mug
\end{lstterm}
which will remove all source code generated by \gum associated with the model \nm. Note that if the user manually alters any of the auto-generated code, the resetting functionality may not work as expected.

\subsection{Input file and node details} \label{sec:gumfile}

\gum files are \YAML files in all but name: they are written in \YAML format and respect \YAML syntax.  The only mandatory nodes for a \gum input file are the \guminline{math} node, specifying details of the \Mathematica package used, and the \guminline{output} node, which selects the \GB backends that \gum should generate code for.

The full set of recognised nodes in a \dgum file is
\begin{description}
  \item \guminline{math}: describes the \Mathematica package used, and subsequently, the model name, plus any other information relating specifically to the package
  \item \guminline{wimp_candidate}: give the PDG code for the annihilating DM candidate in the model
  \item \guminline{decaying_dm_candidate}: give the PDG code for the decaying DM candidate in the model
  \item \guminline{invisibles}: give the PDG codes for particles to be treated as invisibles in collider analyses
  \item \guminline{output}: selects which backends \gum should write output for
  \item \guminline{output_options}: specific options to use for each backend installation.
\end{description}

The \guminline{math} node syntax is
\begin{lstgum}
math:

  # Select the Mathematica package to use,
  # either `feynrules' or 'sarah'
  package: feynrules

  # Choose the name of the model
  model: @\nm@
\end{lstgum}

For specific information on \fr files, see Sec.~\ref{sec:feynrules}, and for \sarah files, see Sec.~\ref{sec:sarah}.

Information about the DM candidate of interest is given by either the \guminline{wimp_candidate} node or the \guminline{decaying_dm_candidate} node.  Note that only one of these nodes can be passed; if both are present in the \dgum file, \gum will throw an error.  Although these nodes are optional, if neither is present, then no output will be written for \darkbit (including the Process Catalogue and direct detection interfaces). The syntax is
\begin{lstgum}
# Select the PDG code of the DM candidate
wimp_candidate: 9900001
\end{lstgum}
in the case of annihilating DM, and similarly
\begin{lstgum}
# Select the PDG code of the DM candidate
decaying_dm_candidate: 9900009
\end{lstgum}
in the case of decaying DM.  Note that only one DM candidate can be specified at present. Future versions of \GB will allow for multiple DM candidates, and for the lightest stable particle (LSP) to be determined by the \cpp{Spectrum} object.

Any additional particles that are to be treated as invisible when calculating missing momentum in collider analyses are given by the \guminline{invisibles} node. They are given in a list, with anti-particle PDG codes also required when necessary. If this node is not present, \colliderbit will use the default list provided in \term{contrib/heputils/include/HEPUtils/Event.h}.

\begin{lstgum}
# Specify the PDGs of any additional invisibles
invisibles: [9900001,-9900001]
\end{lstgum}

The \guminline{output} option specifies for which backends \gum should generate code.
\begin{lstgum}
# Specify outputs: calchep, pythia, spheno,
# vevacious, micromegas
output:
  calchep: true
  spheno: true
  vevacious: false
\end{lstgum}
The default for each possible backend output is \guminline{false}. If the \guminline{output} node is empty, or if all backend output is set to \guminline{false}, \gum will terminate with an error message.

The \guminline{output_options} node allows the user to pass specific settings relevant for each backend to \gum. We briefly go through these in turn. The syntax for this is
\begin{lstgum}
output:
  backend_a: true
  backend_b: true

output_options:
  backend_a:
    # Option given by a single key
    option_a: value_a
  backend_b:
    # Option given by a list
    option_b:
    - entry_1
    - entry_2
    ...
\end{lstgum}

To tell \MG which processes to generate \pythia matrix elements for, the user should provide a list of all BSM processes in \MG syntax under the \guminline{output_options::pythia::collider_processes} sub-node. The user needs to know the names of each particle within \MG in order to fill in this information.

While \pythia is able to perform its own showering for initial jets, these will be very soft. If the user specifically requires hard ISR jets, such as for a monojet signal associated with DM pair production, these matrix elements should be explicitly requested.   In doing so, the user must be careful and aware that collider events are not double counted, i.e. jet matching is performed. We explain our treatment of jet matching in \pythia in Sec.~\ref{sec:pythia}.

For example, to generate matrix elements for monojet and mono-photon production in association with pair production of a DM candidate $X$, one would include
\begin{lstgum}
output:
  pythia: true

output_options:
  pythia:
    # Processes in MadGraph syntax
    collider_processes:
      - p p > ~X ~X
      - p p > ~X ~X j
      - p p > ~X ~X a
\end{lstgum}
The \guminline{collider_processes} sub-node is currently always required if \guminline{pythia:true} is set in the \dgum file.

As with the existing \pythia functionality in \colliderbit, the new function \cpp{getPythia_}\nm\ introduced to \colliderbit by \gum recognises a \YAML option \yaml{pythia_settings}, which can be provided by a user in their GAMBIT input file.  In particular, the boolean sub-option \nm\yaml{:all} of the \yaml{pythia_settings} option allows one to activate all processes specified in the \guminline{collider_processes} entry of the \dgum file.

Other sub-nodes of the \dgum file's \guminline{pythia} entry offer the ability to use the native multiparticle description within \MG (\guminline{multiparticles}), and to select events with the relevant particles in the initial and final state (\guminline{pythia_groups}). An example including all available sub-nodes for the \guminline{pythia} entry is shown below:
\begin{lstgum}
math:
  package: feynrules
  model: newSUSY

output:
  # Generate output for Pythia
  pythia: true

output_options:
  pythia:
    # Define some multiparticles for convenience
    multiparticles:
      - chi0: [chi0_1, chi0_2, chi0_3, chi0_4]
      - chi0bar: [chi0_1, chi0_2, chi0_3, chi0_4]
      - chipls: [chi+_0, chi+_1]
      - chimns: [chi-_0, chi-_1]
    # All processes we want to export to Pythia
    collider_processes:
      - p p grt @~@chi0 chi0bar j
      - p p grt @~@chipls chimns
      - p p grt @~@chimns Hpls
    # Define some groups so we can import processes
    # with these particles in the initial or final
    # states when we do a scan
    pythia_groups:
      - Neutralino: [chi0_1, chi0_2, chi0_3, chi0_4]
\end{lstgum}

In this example, including the example \guminline{pythia_groups} node will generate an additional group of events known as \yaml{newSUSYNeutralino:all}, which can also be set in the \yaml{pythia_settings} option of the new \colliderbit module function \cpp{getPythia_newSUSY}. Setting this flag to \yaml{on} picks out all processes in which any of the particles in the \guminline{pythia_group} is an initial or a final state.  This is useful for when one wishes to simulate events only for a specific subset of the processes for which matrix elements have been generated for the new model.

For \spheno, the user can request to turn loop decays off via the flag \guminline{IncludeLoopDecays},
\begin{lstgum}
math:
  package: sarah
  model: newHDM

output:
  # Generate output for SPheno
  spheno: true

output_options:
  spheno:
    IncludeLoopDecays: false # default: true
\end{lstgum}

\subsection{\fr pathway} \label{sec:feynrules}

Here we describe the process by which \gum can parse a model defined in \fr. For details on how to correctly implement a model in \fr, we refer the reader to the \fr manual~\cite{Alloul:2013bka}. There are many examples of models available on the \fr website.\footnote{\url{http://feynrules.irmp.ucl.ac.be/wiki/ModelDatabaseMainPage}}

\subsubsection{Outputs}

\fr is designed to study particle physics phenomenology at tree level, and does not directly interface to any spectrum generators. \fr is therefore well suited to EFTs and simplified models, as gauge invariance and renormalisability are not typically required in these cases. Because of this, when working from the outputs of \fr, \gum is only able to provide minimal interfaces to the \specbit module and the \GB model database.

\fr is able to output two file formats usable by \gum: \CH (\term{.mdl} files) and \ufo files. \gum uses \term{.mdl} files with \CH to compute tree-level decay rates and DM annihilation cross-sections, and with \mo to compute DM relic densities and direct detection rates. The \ufo files are currently only used by the \MG-\pythia\textsf{8} chain, for collider physics. See Sec.~\ref{sec:code} for details.

\subsubsection{Porting a \fr model to \GB}

To add a model to \GB based upon a \fr file, \gum tries to find the \nm\term{.fr} model file, and any restriction (\term{.rst}) files that the user may wish to include, first in
\begin{lstterm}
gum/contrib/FeynRules/Models/@\nm@/
\end{lstterm}
where the folder \term{gum/} is located inside the \GB root directory. If these model files do not come with the \fr installation, \gum instead looks in
\begin{lstterm}
gum/Models/@\nm@/
\end{lstterm}

To emulate the \fr command
\mathematica{LoadModel["}\nm\mathematica{.fr"}\term{]}
the \dgum file simply needs the entry
\begin{lstgum}
math:
  package: feynrules
  model: @\nm@
\end{lstgum}

Many models hosted on the \fr website and elsewhere often utilise `base' files and extensions, where one model builds upon another. For instance, a model called \yaml{SingletDM} that builds on the Standard Model could be loaded in a \Mathematica session using the \fr command
\mathematica{
LoadModel["SM.fr","SingletDM.fr"].
}
This behaviour is also possible with \gum via the additional option \guminline{base_model}. In this case, \gum expects \term{SM.fr} to be located in \term{Models/SM/} and \term{SingletDM.fr} to be in \term{Models/SingletDM/} (where both paths can be independently relative to \term{gum/} or to \term{gum/contrib/FeynRules/}). A user would indicate this in their input file like so:
\begin{lstgum}
math:
  package: feynrules
  model: SingletDM
  base_model: SM
\end{lstgum}

An additional \fr-only option for the \guminline{math} node includes the ability to load \fr restriction (\term{.rst}) files.  A \fr restriction is useful when considering a restricted subspace of the model at hand. For example, to set the CKM matrix to unity, we can load the restriction file \term{DiagonalCKM.rst} that is shipped with \fr, as
\begin{lstgum}
math:
  ...
  # Specify any restriction files
  restriction: DiagonalCKM
\end{lstgum}
Another \fr-only option is the ability to specify the name of the Lagrangian that \fr should compute the Feynman rules for. The definition of the Lagrangian can either be a single definition from the \fr file:
\begin{lstgum}
math:
  package: feynrules
  model: SingletDM
  base_model: SM
  # Total Lagrangian in SingletDM.fr
  lagrangian: LTotal
\end{lstgum}
or can be given as a string of Lagrangians:
\begin{lstgum}
math:
  package: feynrules
  model: SingletDM
  base_model: SM
  # All symbols defined in SM.fr or SingletDM.fr
  lagrangian: LSM + LDMinteraction + LDMKinetic
\end{lstgum}

After loading the model, \gum performs some diagnostics on the model to ensure its validity, checking that the Lagrangian is Hermitian, and that all kinetic and mass terms are correctly diagonalised according to the \fr conventions. For more details, we refer the reader to the \fr manual~\cite{Alloul:2013bka}.

\subsubsection{Requirements for \fr files} \label{sec:fr_params}

\gum interacts with loaded \fr files via the \mathematica{EParamList} and \mathematica{PartList} commands. To successfully parse the parameter list, every parameter \textit{must} have a \mathematica{BlockName} and \mathematica{OrderBlock} associated with it.

A model implemented in \fr will be parametrised in \GB by the full set of parameters denoted as external, by \mathematica{ParameterType -> External} in the input \term{.fr} file. Additionally, all masses for non-SM particles are added as input parameters, as they are not computed by spectra.

For example, the SM extended by a scalar singlet $S$ via a Higgs portal with the interaction Lagrangian $\mathcal{L} \supset \lambda_{hs} H^\dagger H S^2$ would be parametrised in \GB by the coupling $\lambda_{hs}$, as well as the mass of the new field $m_S$.

The user should not use non-alphanumeric characters (apart from underscores) when defining parameter names (including the \mathematica{ExternalParameter} field), as this will typically result in errors when producing output. The exception to this is a tilde, which is often used to signify a conjugate field, or in the case of \mo a DM candidate.

For the \MG-\pythia pathway to work correctly, each new external parameter must have its \mathematica{InteractionOrder} set. See Sec. 6.1.7 of the \fr manual for details~\cite{Alloul:2013bka}. A fully compliant \fr entry for a parameter looks as follows:
\begin{lstmathematica}
M$Parameters = {
  ...
  gchi == {
    ParameterType    -> External,
    ComplexParameter -> False,
    InteractionOrder -> {NP, 1},
    BlockName        -> DMINT,
    OrderBlock       -> 1,
    Value            -> 1.,
    TeX              -> Subscript[g,\\[Chi]],
    Description      -> "DM-mediator coupling"
  },
  ...
\end{lstmathematica}
where the \mathematica{BlockName}, \mathematica{OrderBlock} and \mathematica{InteractionOrder} are all defined. We also set \mathematica{ComplexParameter} to \mathematica{False}, as \fr is not able to generate \CH files for complex parameters. All parameters that are complex should be redefined as their real and imaginary parts, with all factors of $i$ explicitly placed in the Lagrangian.

For a matrix, the \mathematica{OrderBlock} does not need to be specified,
\begin{lstmathematica}
M$Parameters = {
  ...
  yL == {
    ParameterType    -> External,
    ComplexParameter -> False,
    InteractionOrder -> {BSM, 1},
    Indices          -> {Index[Generation],
                         Index[Generation]},
    BlockName        -> yL,
    Value            -> {yL[1,1]->1, yL[1,2]->0,
                         yL[2,1]->0, yL[2,2]->1},
    TeX              -> Subscript[y,L],
    InteractionOrder -> {NP, 1},
    Description      -> "Left-handed matrix"
  },
  ...
\end{lstmathematica}

In this case, \gum will add 4 model parameters to the model for each matrix index, labelled by \textinline{matrixname_[i]x[j]}, \ie \textinline{yL_1x1, yL_1x2, yL_2x1, yL_2x2} for the above entry. Note that the values for each entry can be set to anything; these will all be set by \GB during a scan.

An example of a particle implementation, for the Majorana DM candidate used in Sec.~\ref{sec:example}, is
\begin{lstmathematica}
M$ClassesDescription = {
  ...
  F[5] == {
    ClassName        -> chi,
    SelfConjugate    -> True,
    Mass             -> {mchi, 1000.},
    Width            -> 0.,
    PDG              -> 52,
    ParticleName     -> "~chi",
  },
  ...
\end{lstmathematica}
Here we see that the \mathematica{ParticleName} begins with a tilde, so that \mo can correctly identify it as a WIMP DM candidate, the PDG code is assigned to 52 (generic spin-1/2 DM, as per the PDG), and the particle mass \mathematica{mchi} will be added as an external parameter. Note that because this particle has \mathematica{SelfConjugate -> True}, \gum does not require the electric charge to be set. If the particle were Dirac, \ie \mathematica{SelfConjugate -> False}, \gum would require the additional entry \mathematica{QuantumNumbers -> \{Q -> 0\}}.

For a particle $\eta$ that should decay, an appropriate entry for the particle width would look like \mathematica{Width -> \{weta, 1.\}}, enabling the contents of the \cpp{DecayTable} to be passed to \CH. Note that in this case, \mathematica{weta} will not be set as a free parameter of the model in \GB, but derived from the model parameters and accessible channels.

Although \fr is able to compute the Feynman rules for a theory containing 4-fermion interactions, it does not support generating \CH files for these models.\footnote{At the time of writing, \textsf{LanHEP} is the only package that supports automatic generation of 4-fermion contact interactions for \CH files.} The first release of \gum does not support theories implemented in \fr with 4-fermion interactions, however such support is planned for future releases.

\subsection{\sarah pathway} \label{sec:sarah}

\subsubsection{Outputs}

As shown in Table~\ref{tab::outputs}, \sarah is able to generate output for \CH, \mo, \pythia, \spheno and \veva. As \sarah is able to generate \CH, \MG/\pythia and \mo output, it can mirror the capabilities of \fr in the context of \gum.

\sarah has been labeled a `spectrum generator generator', as it can also automatically write \Fortran source code for \spheno for a given model. \gum is able to automatically patch the \spheno source code generated by \sarah, and write a frontend interface to that \sarah-\spheno version.

\subsubsection{Porting a \sarah model to \GB}

To add a model to \GB based upon a \sarah file, the model file \nm\term{.m} must be located in
\begin{lstterm}
gum/contrib/SARAH/Models/@\nm@/
\end{lstterm}
or
\begin{lstterm}
gum/Models/@\nm@/
\end{lstterm}
The usual \sarah files \term{parameters.m} and \term{particles.m} should also be present in one of these locations.  To generate spectra via \spheno, a \term{SPheno.m} file must also be provided in the same directory.

\gum loads a new model in \sarah by invoking the command \mathematica{Start[\"}\nm\mathematica{\"}\term{]}, which is selected by the \dgum entries

\begin{lstgum}
math:
  package: sarah
  model: @\nm@
\end{lstgum}

In order to validate the model \gum uses the \sarah command \mathematica{CheckModel[]}. \sarah provides the results of the \mathematica{CheckModel[]} function only to \term{stdout} and via error messages. \gum therefore captures the output and message streams from \Mathematica in order to gather this information, and decides whether the errors should be considered fatal or not. Non-fatal errors, including gauge anomalies, possible allowed terms in the Lagrangian or missing Dirac spinor definitions, are directed to \gum's own standard output as warnings. Fatal errors, such as non-conservation of symmetries or those associated with particle and parameter definitions, cause \gum to abort, as subsequent steps are guaranteed to fail in these cases.

\subsubsection{Requirements for \sarah files} \label{sec:sarah_params}

As with \fr, \gum extracts information from \sarah about the parameters and particles in the model. These are collected by \sarah in the \mathematica{ParameterDefinitions} and \mathematica{ParticleDefinitions} lists, respectively.

Definitions for new model parameters are located in the \term{parameters.m} file within the \sarah model folder. A well-defined entry for a new \sarah parameter looks as follows:
\begin{lstmathematica}
ParameterDefinitions = {
  ...
  {gchi,
    {
      Description -> "DM-mediator coupling",
      LesHouches  -> {DMINT, 1},
      OutputName  -> "gchi",
      LaTeX       -> "g_\\chi"
    }
  },
  ...
}
\end{lstmathematica}
where the \mathematica{LesHouches} block and respective index are required fields.

For a matrix, the index does not need to be specified:
\begin{lstmathematica}
ParameterDefinitions = {
  ...
  {YN,
    {
      Description -> "Yukawa for N field",
      LesHouches  -> YN,
      LaTeX       -> "Y_{\\rm N}",
      OutputName  -> yn
    },
  },
  ...
}
\end{lstmathematica}
This instructs \gum to add a \mathematica{LesHouches} block \mathematica{YN} to the \cpp{SimpleSpec} definition, which will be filled by a spectrum generator.

\gum is concerned with the properties of physical particles in the mass basis (designated \mathematica{[EWSB]} in \sarah). An example particle implementation from the \term{particles.m} file is:
\begin{lstmathematica}
ParticleDefinitions[EWSB] = {
  ...
  {ss,
    {
      Description    -> "Scalar singlet",
      Mass           -> LesHouches,
      PDG            -> {51},
      ElectricCharge -> 0,
      OutputName     -> "~Ss",
      LaTeX          -> "S"
    }
  },
  ...
}
\end{lstmathematica}
Here the important entries are\begin{itemize}
\item the \mathematica{Mass} entry, where \mathematica{Mass -> LesHouches} signifies that the particle mass will be provided by the \GB\ \cpp{Spectrum} object (whether that is filled using \spheno or a tree level calculation),
\item the \mathematica{PDG} entry, which specifies a list over all generations for the mass eigenstates (in this example there is just one), and
\item the \mathematica{ElectricCharge} field.
\end{itemize}

Note that \sarah has default definitions for many particles and parameters in \metavar{SARAH\_dir}\term{/Models/particles.m} and \metavar{SARAH\_dir}\term{/Models/parameters.m}.  Their properties can be inherited, or overwritten, via the \mathematica{Description} field.

Information about mixing matrices is stored by \sarah in the variable \mathematica{DEFINITION[EWSB][MatterSector]}. From this variable \gum learns the names of the mixing matrices associated with each particle. For Weyl fermions, \gum requests the name of the associated Dirac fermion, stored in the variable \mathematica{DEFINITION[EWSB][MatterSector]}. As an example, the mixing matrices for the electroweakino sector of the MSSM are extracted as
\begin{lstmathematica}
 DEFINITION[EWSB][MatterSector][[;;,2]] = {
 ...
   {L0, ZN},
   {{Lm, UM}, {Lp, UP}}
 ...
 }
\end{lstmathematica}
which associates the matrix \mathematica{ZN} with the Weyl fermion \mathematica{L0} (neutralinos) and the matrixes \mathematica{UM} and \mathematica{UP} with \mathematica{Lm} (negative charginos) and \mathematica{Lp} (positive charginos). As these are Weyl fermions, the Dirac eigenstates are
\begin{lstmathematica}
 DEFINITION[EWSB][DiracSpinors] = {
 ...
   Chi -> {L0, conj[L0]},
   Cha -> {Lm, conj[Lp]}
 ...
 }
\end{lstmathematica}
\gum thus knows to assign the mixing matrix \mathematica{ZN} to Dirac-eigenstate neutralinos \mathematica{Chi}, as well as the matrices \mathematica{UM} and \mathematica{UP} to Dirac-eigenstate charginos \mathematica{Cha}.

As opposed to \fr, where all parameters and particle masses become \GB model parameters, the \sarah pathway attempts to optimise this list through various means. In the absence of a spectrum generator (e.g.\ \spheno, see below), almost all the parameters in \mathematica{ParameterDefinitions} become model parameters.  Only those with explicit dependencies on other parameters are removed, i.e.\ those with the \mathematica{Dependence} or \mathematica{DependenceSpheno} fields. In addition, \sarah provides tree-level relations for all masses, via \mathematica{TreeMass[}\pn\mathematica{,EWSB]}, so even in the absence of a spectrum generator, none of the particle masses become explicit model parameters. For models with BSM states that mix together into mass eigenstates\footnote{Technically this is done by checking if the PDG list for any of the BSM particles contains more than one entry.}, the tree-level masses are not used and an error is thrown to inform the user of the need to use a spectrum generator.

If the user elects in their \dgum file to generate any outputs from \sarah for specific backends, \gum requests that \sarah generate the respective code using the relevant \sarah commands.  These are \mathematica{MakeCHep[]} for \CH and \mo, \mathematica{MakeUFO[]} for \MG/\pythia, \mathematica{MakeSPheno[]} for \spheno and \mathematica{MakeVevacious[]} for \veva.

When \spheno output is requested, \gum interacts further with \sarah in order to obtain all necessary information for spectrum generation:
\begin{enumerate}
 \item Replace parameters and masses with those in \spheno. The parameter names are obtained using the \mathematica{SPhenoForm} function operating on the lists \mathematica{listAllParametersAndVEVs} and \mathematica{NewMassParameters}. The particle masses are obtained just by using the \mathematica{SPhenoMass[}\pn\mathematica{]} command.

 \item Extract the names and default values of the parameters in the \mathematica{MINPAR} and \mathematica{EXTPAR} blocks, as defined in the model file \term{SPheno.m}. For each of these, store the boundary conditions, also from \term{SPheno.m}, that match the \mathematica{MINPAR} and \mathematica{EXTPAR} parameters to those in the parameter list. Note that as of \gum \textsf{1.0}, only the boundary conditions in \mathematica{BoundaryLowScaleInput} are parsed.

 \item Remove from the parameter list those parameters that will be fixed by the tadpole equations, as they are not free parameters. These are collected from the list \mathematica{ParametersToSolveTadpoles} as defined in \term{SPheno.m}.

 \item Get the names of the blocks, entries and parameter names for all SLHA input blocks ending in \mathematica{IN}, e.g.\ \mathematica{HMIXIN}, \mathematica{DSQIN}, etc. \sarah provides this information in the list \mathematica{CombindedBlocks}.

 \item Register the values of various flags needed to properly set up the interface to \spheno.  These are \term{"SupersymmetricModel"}, \term{"OnlyLowEnergySPheno"}, \term{"UseHiggs2LoopMSSM"} and \term{"SA`AddOneLoopDecay"}.

\end{enumerate}

\section{A worked example}\label{sec:example}

To demonstrate the process of adding a new model to \GB with \gum, in this section we provide a simple worked example.  Here we use \gum to add a model to \GB, perform a parameter scan, and plot the results with \pippi~\cite{pippi}. This example is designed with ease of use in mind, and can be performed on a personal computer in a reasonable amount of time. For this reason we select a simplified DM model, implemented in \fr.

In this example, we consider constraints from the relic density of dark matter, gamma-ray indirect detection and traditional high-mass direct detection searches.  It should be noted that this is an example, not a full global scan, so we do not use all of the information available to us -- a real global fit of this model would consider nuisance parameters relevant to DM, as well as a full set of complementary likelihoods such as from other indirect DM searches, low-mass direct detection searches, and cosmology.

The \fr model file, \dgum file, \gambit input file and \term{pip} file used in this example can be found within the \term{Tutorial} folder in \gum.

\subsection{The model}

The model is a simplified DM model, where the Standard Model is extended by a Majorana fermion $\chi$ acting as DM, and a scalar mediator $Y$ with a Yukawa-type coupling to all SM fermions, in order to adhere to minimal flavour violation. The DM particle is kept stable by a $\mathcal{Z}_2$ symmetry under which it is odd, $\chi \rightarrow -\chi$, and all other particles are even. Both $\chi$ and $Y$ are singlets under the SM gauge group.

Here, we assume that any mixing between $Y$ and the SM Higgs is small and can be neglected. This model has been previously considered in e.g.~\cite{Buckley:2014fba,Arcadi:2017kky} and is also one of the benchmark simplified models used in LHC searches~\cite{Abdallah:2015ter,Sirunyan:2019gfm,Aaboud:2019yqu}. The model Lagrangian is
\begin{align}
  \mathcal{L} &= \mathcal{L}_\mathrm{SM} + \frac{1}{2}\overline{\chi}\left(i\slashed{\partial}-m_\chi\right)\chi +\frac{1}{2}\partial_\mu Y \partial^\mu Y - \frac{1}{2} m_Y^2 Y^2 \nonumber\\
              &- \frac{g_\chi}{2} \overline{\chi}\chi Y -\frac{c_Y}{2} \sum_f y_f \overline{f} f Y \,.
\end{align}

Note that this theory is not $SU(2)_L$ invariant.  One possibility for a `realistic' model involves $Y$-Higgs mixing, which justifies choosing the  $Y\overline{f}f$ couplings to be proportional to the SM Yukawas $y_f$.

The free parameters of the model are simply the dark sector masses and couplings, $\{m_\chi$, $m_Y$, $c_Y$, $g_\chi\}$. In this example we follow the \fr pathway, working at tree level.

\subsection{The .gum file}

Firstly, we need to add the \fr model file to the \gum directory. The model is named `MDMSM' (Majorana DM, scalar mediator). Starting in the \gum root directory, we first create the directory that the model will live in, and move the example file from the \term{Tutorial} folder to the \gum directory:
\begin{lstterm}
mkdir Models/MDMSM
cp Tutorial/MDMSM.fr Models/MDMSM/
\end{lstterm}

As we are working with \fr, the only backends that we are able to create output for are \CH, \mo and \MG/\pythia. For the sake of speed, in this tutorial we will not include any constraints from collider physics.  This is also a reasonable approximation, as for the mass range that we consider here, the constraints from e.g.\ monojet, dijet and dilepton searches are subleading (see e.g.\ Ref.\ \cite{Buckley:2014fba} and Appendix \ref{app:collider_validation}).  We therefore set \guminline{pythia: false}. The contents of the supplied \dgum file are simple:
\begin{lstgum}
math:
  # Choose FeynRules
  package: feynrules
  # Name of the model
  model: MDMSM
  # Model builds on the Standard Model FeynRules file
  base_model: SM
  # The Lagrangian is defined by the DM sector (LDM),
  # defined in MDMSM.fr, plus the SM Lagrangian (LSM)
  # imported from the `base model', SM.fr
  Lagrangian: LDM + LSM
  # Make CKM matrix = identity to simplify output
  restriction: DiagonalCKM

# PDG code of the annihilating DM candidate
# in the FeynRules file
wimp_candidate: 52

# Select outputs for DM physics.
# Collider physics is not as important in this model.
output:
  pythia: false
  calchep: true
  micromegas: true
\end{lstgum}
Note the selection of the PDG code of the DM particle as 52, so that if we were to use \pythia, $\chi$ would be correctly identified as invisible.

We can run this from the \gum directory,
\begin{lstterm}
./gum -f Tutorial/MDMSM@.gum@
\end{lstterm}
and \gum will automatically create all code needed to perform a fit using \GB. On a laptop with an Intel Core i5 processor, \gum takes about a minute to run. All that remains now is to (re)compile the relevant backends and \GB, and the new model will be fully implemented, and ready to scan. \gum prints a set of suggested build commands to standard output to build the new backends and \GB itself.  Starting from the \textinline{gum} directory, these are
\begin{lstterm}
cd ../build
cmake ..
make micromegas_MDMSM
make calchep
make -j@\metavar{n}@ gambit
\end{lstterm}
where \metavar{n} specifies the number of processes to use when building.

Note that \gum does not adjust any \cmake flags used in previous \GB compilations, so the above commands assume that the user has already configured \GB appropriately and built any relevant samplers before running \gum.  A user wishing to instead configure and build \GB from scratch after running \gum, in order to e.g.\ run the example scan of Sec.\ \ref{sec:example} using differential evolution sampling and \mpi parallelisation, would need to instead do (again, starting from the \textinline{gum} directory)
\begin{lstterm}
cd ../build
cmake -D WITH_MPI=ON ..
make diver
cmake ..
make micromegas_MDMSM
make calchep
make gamlike
make ddcalc
make -j@\metavar{n}@ gambit
\end{lstterm}
For more thorough \cmake instructions, see the \term{README} in the \term{gum/Tutorial}, and \term{CMAKE_FLAGS.md} in the \GB root directory.

\subsection{Phenomenology and Constraints}
\label{sec:pheno}

The constraints that we will consider for this model are entirely in the DM sector, as those from colliders are less severe. (Collider constraints are investigated in Appendix A.) The dark matter constraints are:

\begin{itemize}
  \item Relic abundance: computed by \mo, and employed as an upper bound, in the spirit of effective DM models.
  \item Direct detection: rates computed by \mo, likelihoods from XENON1T 2018~\cite{Aprile:2018dbl} and LUX 2016~\cite{LUX2016}, as computed with \ddcalc \cite{DarkBit,HP,SSDM2}.
  \item Indirect detection: \emph{Fermi}-LAT constraints from gamma-ray observations of dwarf spheroidal galaxies (dSphs)~\cite{LATdwarfP8}. Tree level cross-sections are computed by \CH, $\gamma$ ray yields are consequently computed via \darksusy \cite{darksusy4,darksusy}, and the constraints are applied by \gamlike \cite{DarkBit}.
\end{itemize}

As the relic density constraint is imposed only as an upper bound, we rescale all DM observables by the fraction of DM, $f=\Omega_\chi/\Omega_\textrm{DM}$.

Here we will use the \GB input file \term{gum/} \term{Tutorial/MDMSM_Tute.yaml}. Although it does contain a little more than the \GB input file automatically generated by \gum (\term{yaml_files/MDMSM_example.yaml}), it is still fairly standard, so we will cover only the important sections here. For an overview of \YAML files in \GB, we refer the reader to Sec. 6 of the \GB manual~\cite{gambit}.

Firstly the parameters section indicates
all models required for this scan: not just the MDMSM parameters, but also SM parameters, nuclear matrix elements and DM halo parameters. The parameter range of interest for the MDMSM model will be masses ranging from $45$\,GeV to $10$\,TeV, and dimensionless couplings ranging from $10^{-4}$ to $4\pi$.  We will scan each of these four parameters logarithmically.
\begin{lstyaml}
Parameters:

  # Import some default GAMBIT SM values
  StandardModel_SLHA2: !import
   @\yamlvalue{include/StandardModel\_SLHA2\_defaults.yaml}@

  # Higgs sector is defined separately in GAMBIT
  StandardModel_Higgs:
    mH: 125.09

  # Our dark matter model, implemented by GUM
  MDMSM:
    mchi:
      range: [45@\yamlvalue{, 10000}@]
      prior_type: log
    mY:
      range: [45@\yamlvalue{, 10000}@]
      prior_type: log
    gchi:
      range: [1e-4@\yamlvalue{, 12.566}@]
      prior_type: log
    cY:
      range: [1e-4@\yamlvalue{, 12.566}@]
      prior_type: log

  # Default halo parameters for the example
  Halo_gNFW_rho0:
    rho0: 0.3
    v0: 240
    vesc: 533
    vrot: 240
    rs: 20.0
    r_sun: 8.5
    alpha: 1
    beta: 3
    gamma: 1

  # Nuclear matrix parameters, also default
  nuclear_params_sigmas_sigmal:
    sigmas: 43
    sigmal: 58
    deltau: 0.842
    deltad: -0.427
    deltas: -0.085
\end{lstyaml}
The \yaml{ObsLikes} section includes likelihoods concerning the relic density, indirect detection from dSphs, and direct detection experiments.
\begin{lstyaml}
ObsLikes:
    # Relic density
    - capability: lnL_oh2
      purpose:    LogLike

    # Indirect detection
    - capability: lnL_FermiLATdwarfs
      purpose:    LogLike

    # Direct detection: LUX experiment
    - capability: LUX_2016_LogLikelihood
      purpose:    LogLike

    # Direct detection: XENON1T experiment
    - capability: XENON1T_2018_LogLikelihood
      purpose:    LogLike
\end{lstyaml}
The \yaml{Rules} section uniquely specifies the functions to use for the dependency resolver:
\begin{lstyaml}
Rules:
  # Use MicrOmegas to compute the relic density
  - capability: RD_oh2
    function: RD_oh2_MicrOmegas

  # Choose to implement the relic density
  # likelihood as an upper bound, not a detection
  - capability: lnL_oh2
    function: lnL_oh2_upperlimit

  # Choose to use detailed Fermi Pass 8 dwarf
  # likelihood from gamlike
  - capability: lnL_FermiLATdwarfs
    function:  lnL_FermiLATdwarfs_gamLike

  # Choose to get decays from regular DecayBit
  # function, not from an SLHA file nor SPheno.
  - capability: decay_rates
    function: all_decays

  # Choose to rescale signals in direct and indirect
  # detection by the relic density fraction
  - capability: RD_fraction
    function: RD_fraction_leq_one
\end{lstyaml}

The scanner section selects the differential evolution sampler \diver \cite{ScannerBit} with a fairly loose stopping tolerance of $10^{-3}$ and a working population of 10,000 points.
\begin{lstyaml}
Scanner:

  # Select differential evolution (DE) scanner
  use_scanner: de

  scanners:

    # Select settings for DE with Diver
    de:
      plugin: diver
      like: LogLike
      NP: 10000
      convthresh: 1e-3
      verbosity: 1
\end{lstyaml}
To perform the scan we copy the \GB input file to the \term{yaml\_files} folder within the \GB root directory. This is a necessary step, as we need to \yaml{\!import} the appropriate Standard Model \YAML file from the relative path \term{include} (i.e.\ the folder \term{yaml_files/include} in the \GB root directory).  From the \GB root directory, we
\begin{lstterm}
cp gum/Tutorial/MDMSM_Tute.yaml yaml_files/
\end{lstterm}
and run \GB with \metavar{n} processes,
\begin{lstterm}
mpirun -n @\metavar{n}@ gambit -f yaml_files/MDMSM_Tute.yaml
\end{lstterm}

The above scan should converge in a reasonable time on a modern personal computer; this took 11\,hr to run using 4 cores on a laptop with an i5-6200U CPU @ 2.30GHz, sampling 292k points in total.  The results of this scan are shown below.

Note that whilst the scan has converged statistically, the convergence criterion that we set in the input file above is not particularly stringent, so many of the contours presented in this section are not sampled well enough to be clearly defined. A serious production scan would typically be run for longer, and more effort made to map the likelihood contours more finely. Nonetheless, the samples generated are more than sufficient to extract meaningful physics.

Once the scan has finished, we can plot the result using \pippi~\cite{pippi}. As \diver aims to finds the maximum likelihood point, we will perform a profile likelihood analysis with \pippi. Assuming that \pippi is in \term{\$PATH}, do
\begin{lstterm}
cd gum/Tutorial
pippi MDMSM.pip
\end{lstterm}
which will produce plots of the four model parameters against one another, as well against as a raft of observables such as the relic abundance and spin-independent cross-section (rescaled by $f$).

\begin{figure}[tpb]
  \centering
  \includegraphics[height=0.84\columnwidth]{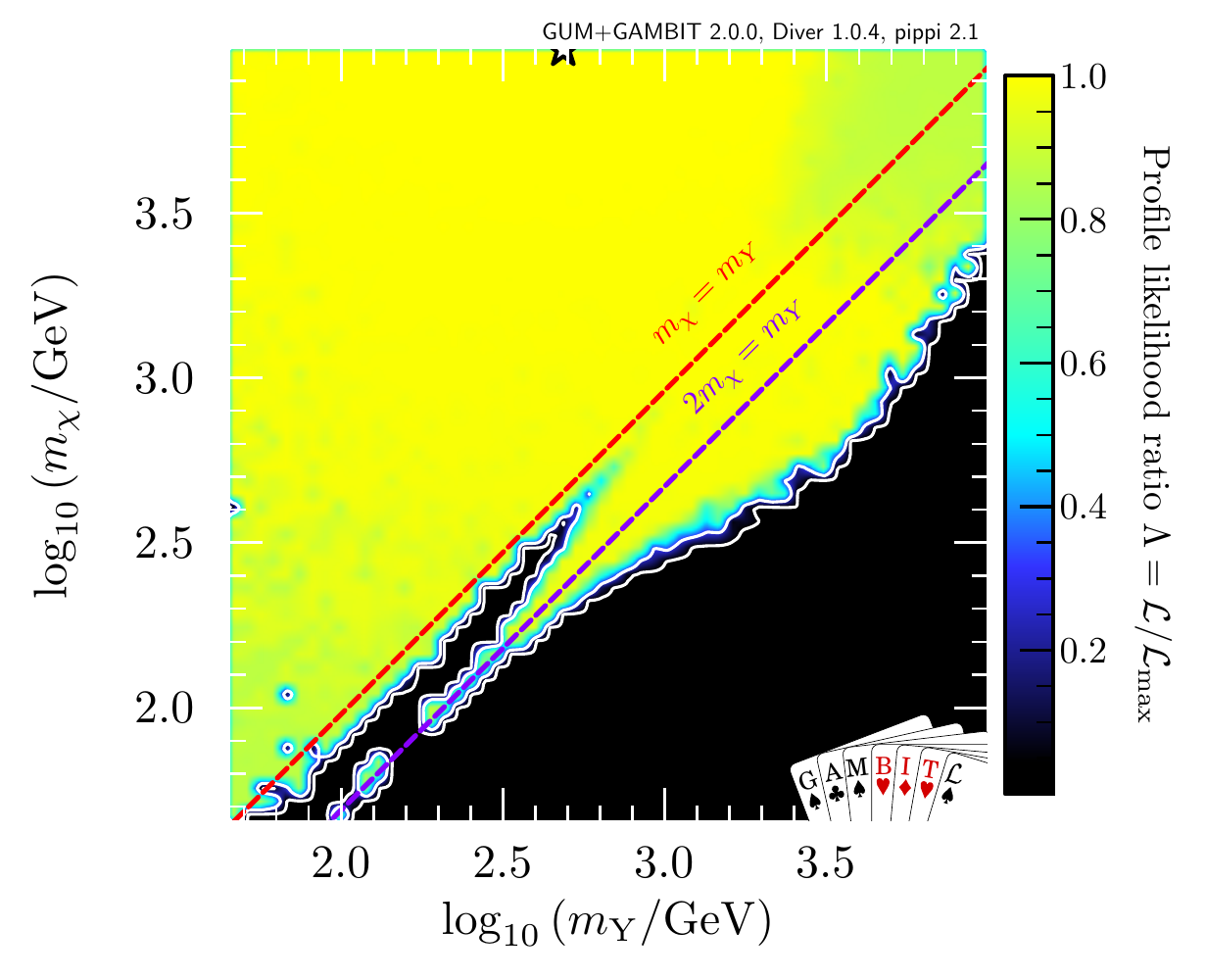}
  \includegraphics[height=0.84\columnwidth]{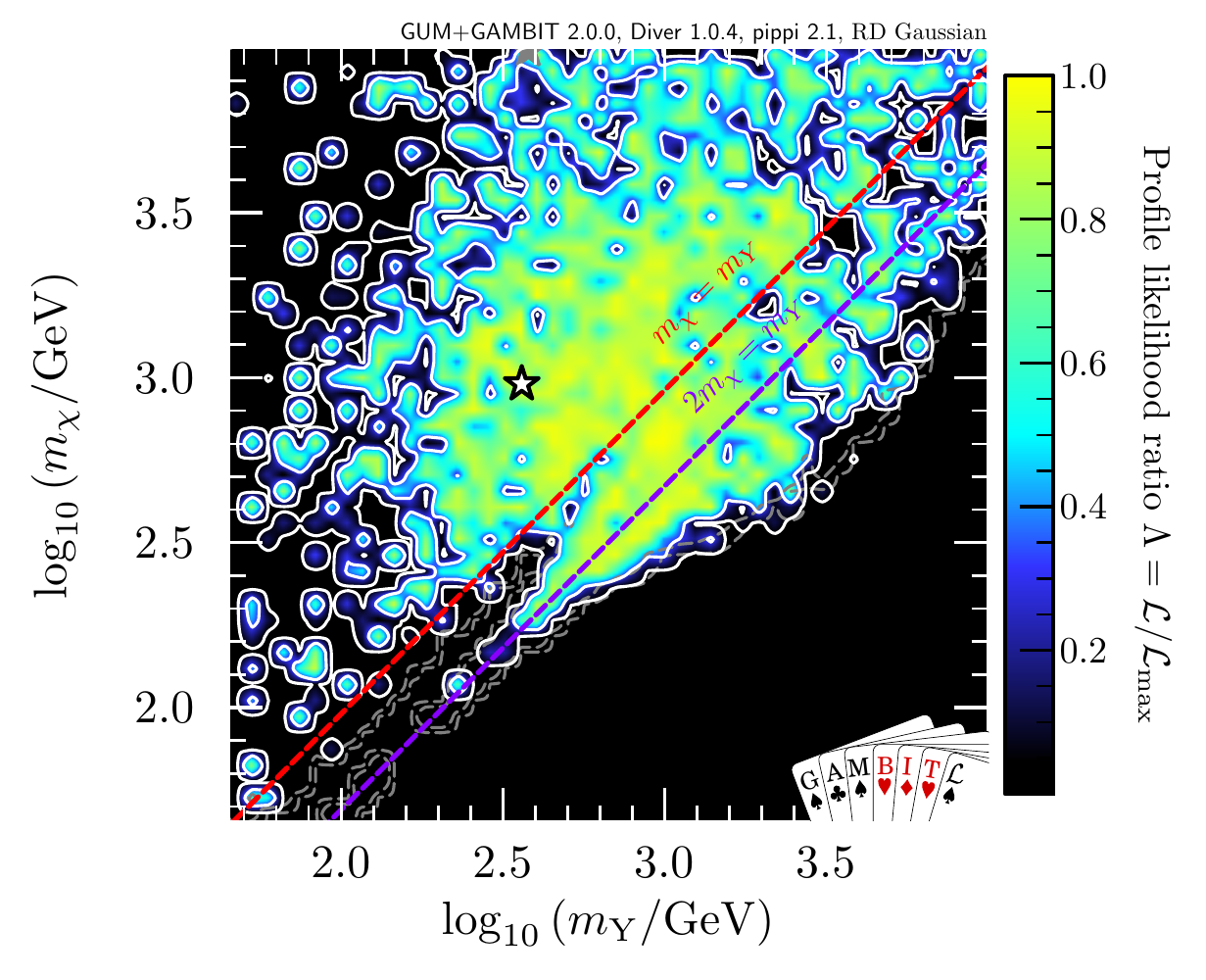}
  \vspace{-2mm}
  \caption{Profile likelihood in the $m_\chi$--$m_Y$ plane with the relic density as an upper bound (upper panel) and as an observation (lower panel).   Above the red dashed line at $m_\chi=m_Y$, DM can annihilate into $Y$ bosons.  The purple dashed line at $2m_\chi=m_Y$ indicates the region where DM can annihilate on resonance.  Contour lines show the $1$ and $2\sigma$ confidence regions. The white star shows the best-fit point. The grey contours in the lower panel the $1$ and $2\sigma$ contours from the upper panel.}
  \label{fig::mchi_vs_mY}
\end{figure}

\subsection{Results}

The upper panel of Fig.~\ref{fig::mchi_vs_mY} shows the profile likelihood in the plane of the DM mass $m_\chi$ against the mediator mass $m_Y$. The relic density requirement maps out the structure in the same plane.  There are two sets of solutions: firstly when the DM is heavier than the mediator, $m_\chi > m_Y$ (bordered by the red dashed line in Fig.~\ref{fig::mchi_vs_mY}), and secondly where DM annihilates on resonance, $2m_\chi \approx m_Y$ (centred on the purple dashed line in Fig.~\ref{fig::mchi_vs_mY}).

When $m_\chi < m_Y$ and the $YY$ annihilation channel is not kinematically accessible, annihilation predominantly occurs via an $s$-channel $Y$ to $b\overline{b}$ or $t\overline{t}$, depending on the DM mass.  In this case, the only way to efficiently deplete DM in the early Universe is when annihilation is on resonance, $m_\chi \approx m_Y/2$. Away from the resonance when the $YY$ channel is closed, even couplings of $4\pi$ are not large enough to produce a sufficiently high annihilation cross-section to deplete the thermal population of $\chi$ to below the observed value.

When kinematically allowed, $\chi\overline{\chi} \rightarrow Y \rightarrow t\overline{t}$ is the dominant process responsible for depleting the DM abundance in the early Universe. When $m_\chi < m_t$ and $m_\chi < m_Y$, the only way to produce the correct relic abundance is when exactly on resonance, $2m_\chi = m_Y$, annihilating mostly to $b\overline{b}$. The effect of the $t$ threshold can clearly be seen in Fig.~\ref{fig::mchi_vs_mY}: as the $\chi\overline{\chi}\rightarrow t\overline{t}$ channel opens up, the contours do not trace the resonance $2m_\chi = m_Y$ quite as tightly. This is because the cross-section to $\overline{t}t$ is significantly larger, as the mediator coupling to the SM leptons are proportional to their Yukawas. This means that near the resonance region it is far easier to satisfy the DM abundance constraint, which leads to the spread about the purple line for $m_\chi > m_t$ in Fig.~\ref{fig::mchi_vs_mY}.

\begin{figure}[tbp]
  \centering
  \includegraphics[height=0.84\columnwidth]{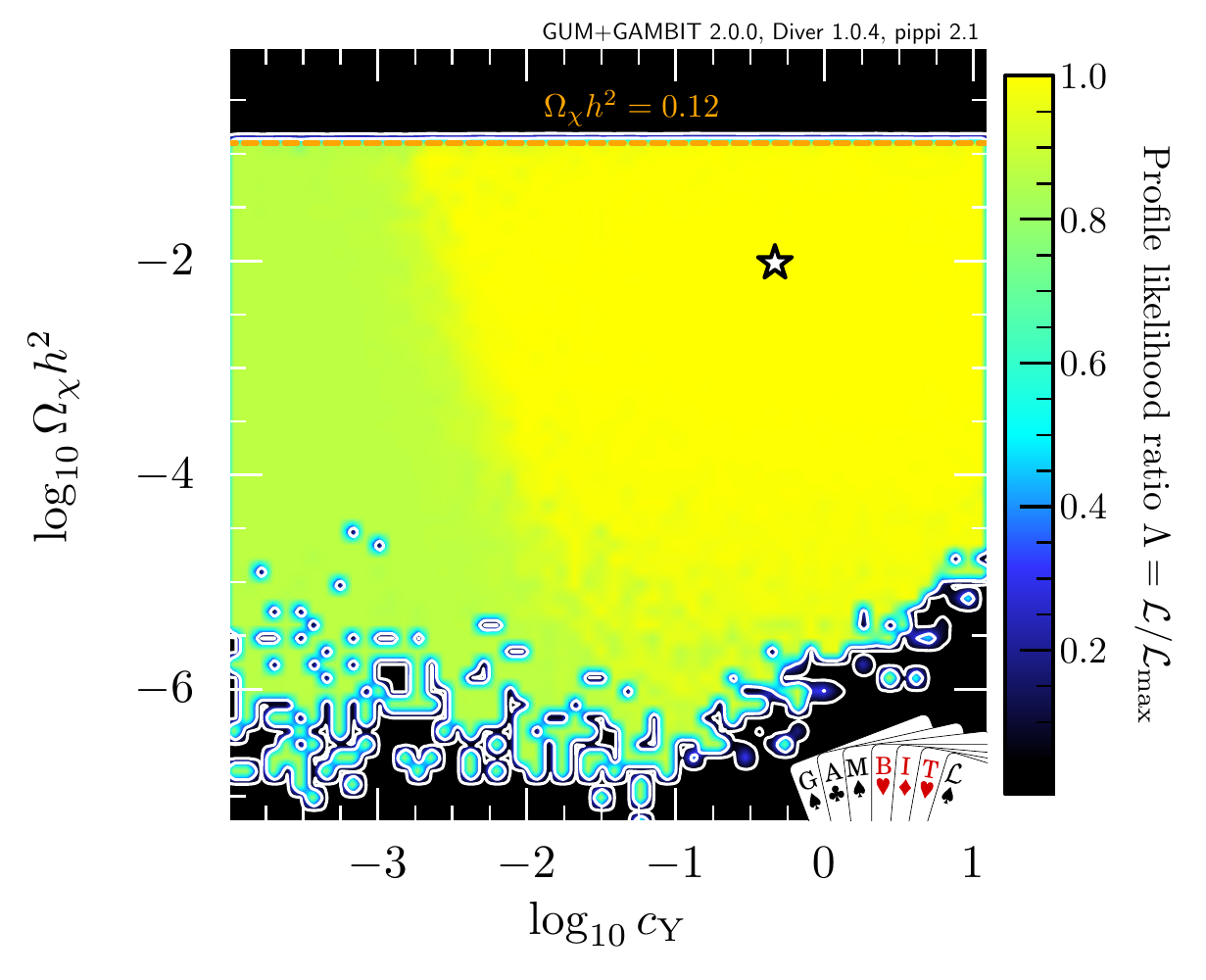}
  \includegraphics[height=0.84\columnwidth]{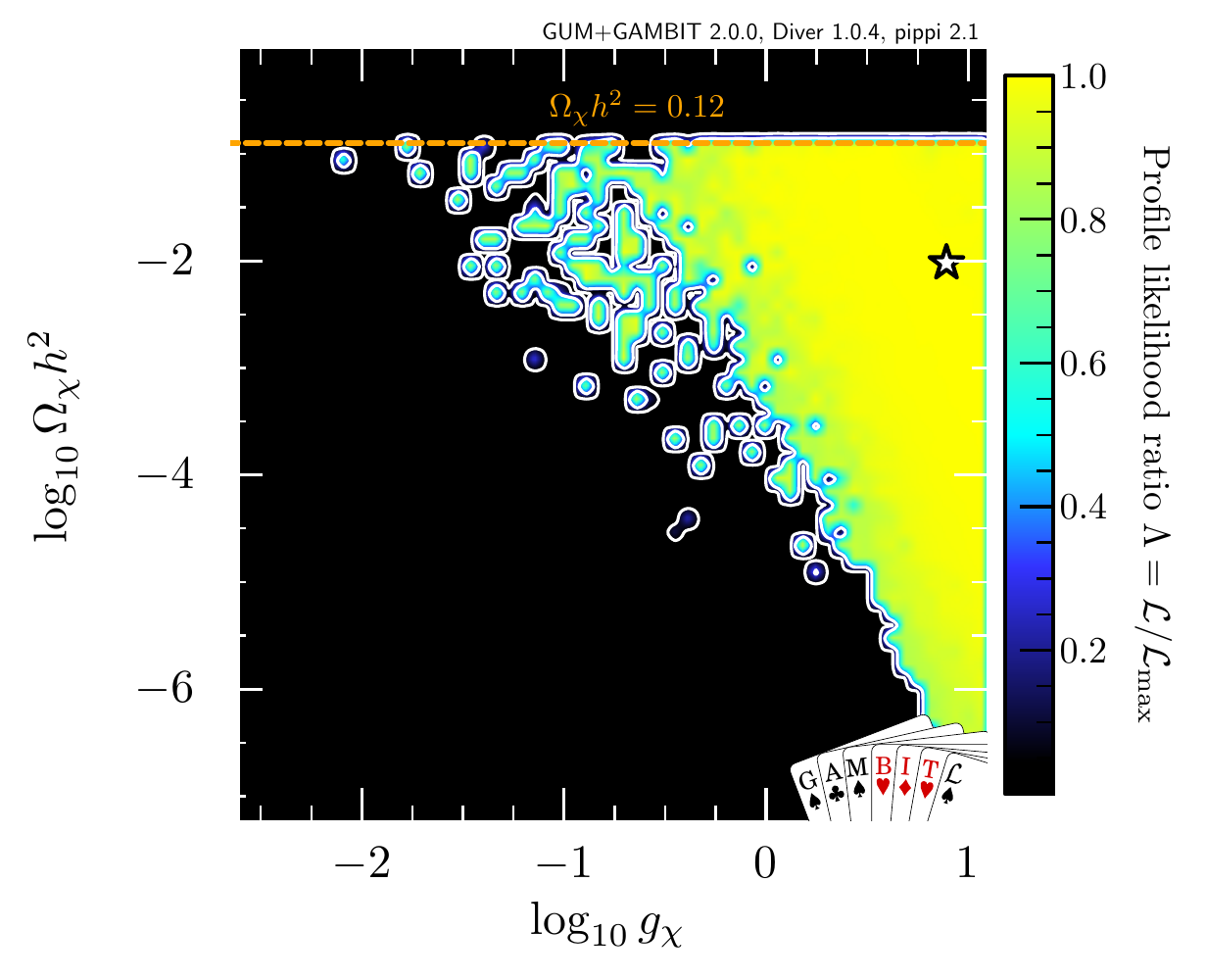}
  \includegraphics[height=0.84\columnwidth]{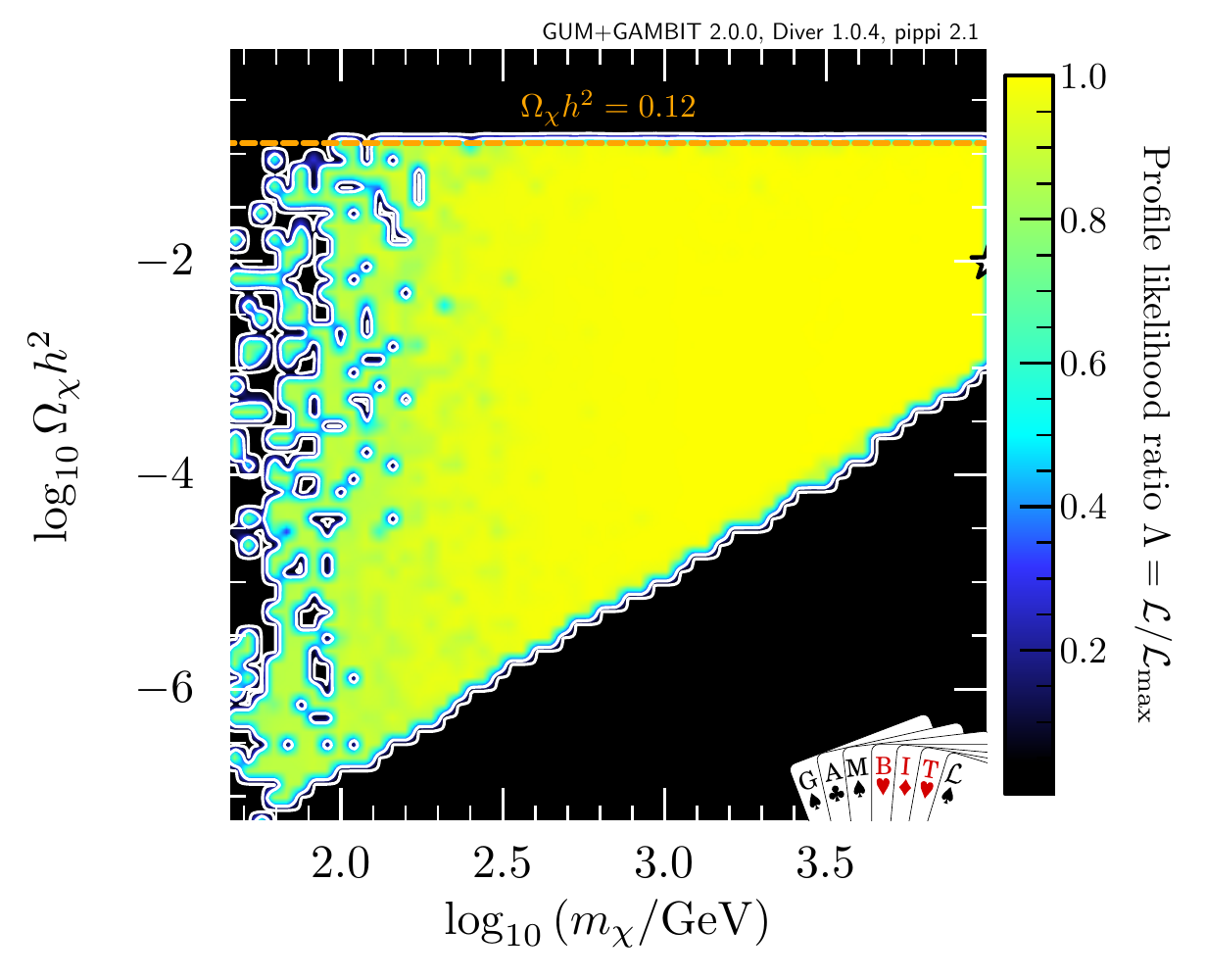}
  \vspace{-2mm}
  \caption{Profile likelihood in the $\Omega_\chi h^2$--$c_Y$, (top) $\Omega_\chi h^2$--$g_\chi$ (centre), and $\Omega_\chi h^2$--$m_\chi$ (bottom) planes. The orange dashed line shows the standard $\Lambda$CDM limit from \emph{Planck} \cite{Ade:2015xua}. Contour lines show $1$ and $2\sigma$ confidence regions, and the white star the best-fit point.}
  \label{fig::relic_densities}
\end{figure}

\begin{figure}
  \centering
  \includegraphics[height=0.84\columnwidth]{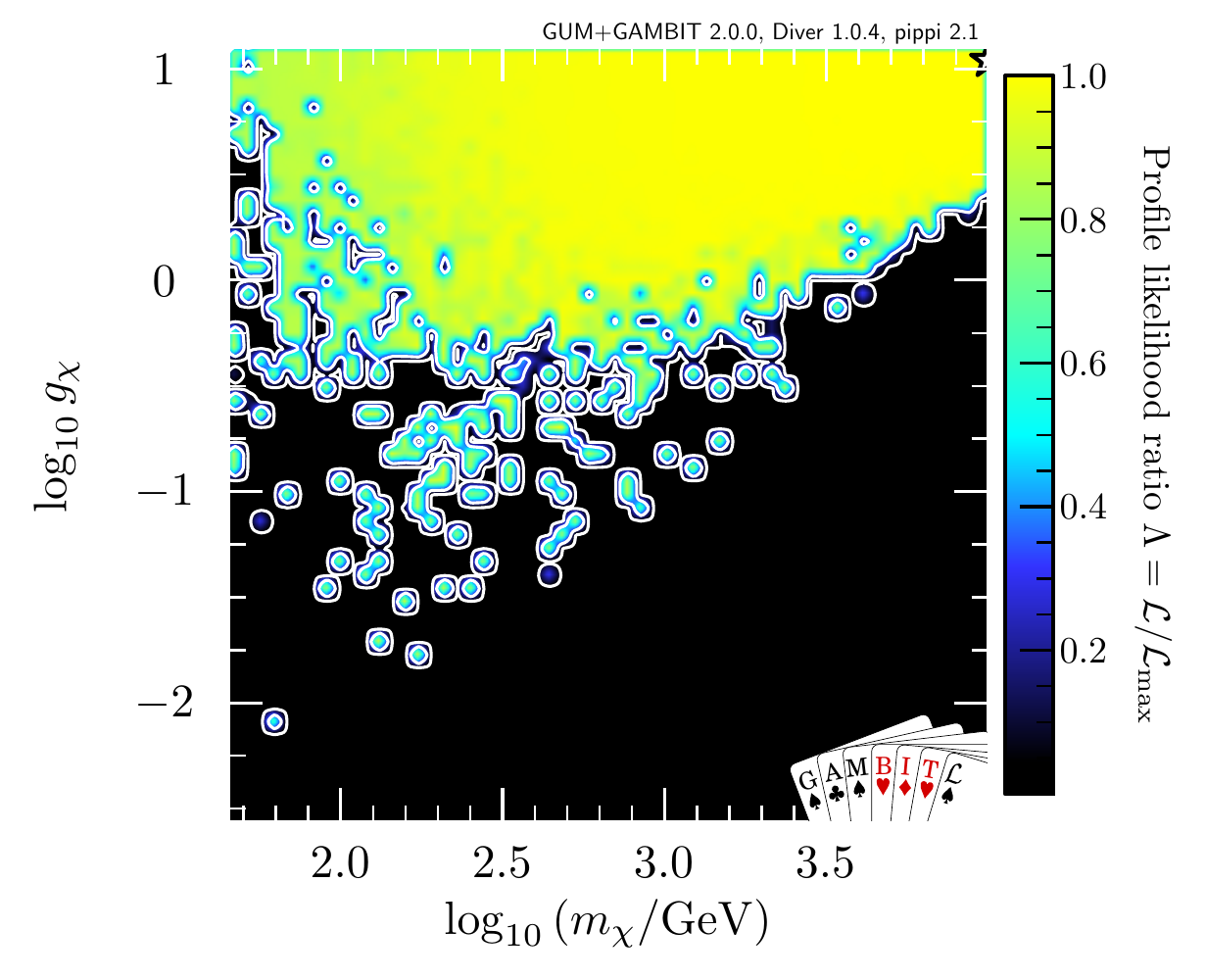}
  \includegraphics[height=0.84\columnwidth]{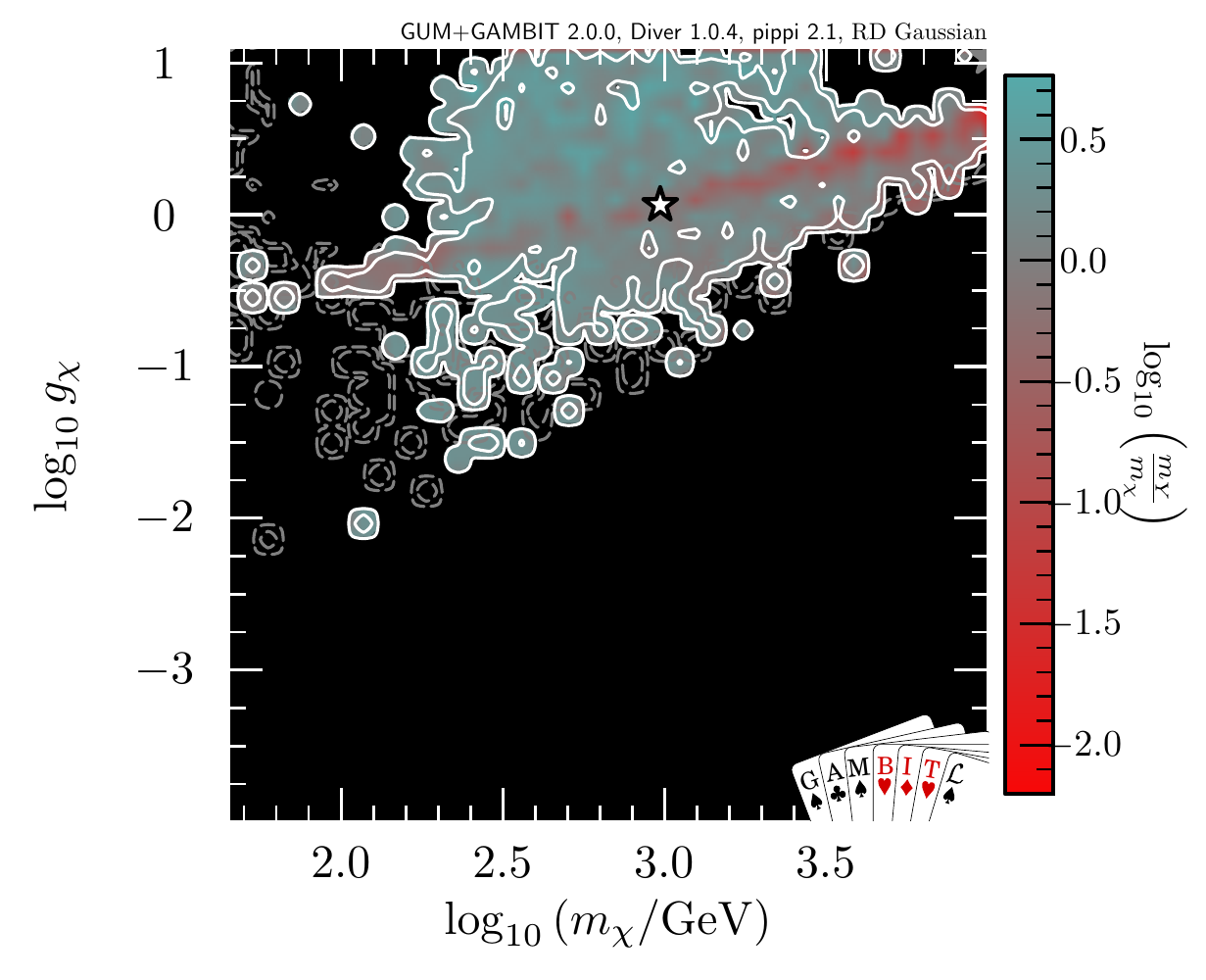}
  \caption{Upper panel: profile likelihood in the $m_\chi$--$g_\chi$ plane when the relic density is an upper bound. Lower panel: the $m_\chi$--$g_\chi$ plane coloured by $m_Y/m_\chi$, when treating the relic density as an observation. Shading and white contour lines represent $1$ and $2\sigma$ confidence regions. Grey contours in the lower panel are the $1$ and $2\sigma$ regions from the upper panel, where the relic density is used as an upper bound. The white star corresponds to the best-fit point.}
  \vspace{-2mm}
  \label{fig::mchi_vs_gchi}
\end{figure}

When the DM candidate is heavier than the mediator, the process $\chi\overline{\chi} \rightarrow Y Y$ is kinematically accessible, and proceeds via $t$-channel $\chi$ exchange. When this channel is open, the correct relic abundance, $\Omega_\chi h^2$, can be acquired independently of $c_Y$ by adjusting $m_\chi$ and $g_\chi$. This can be seen in Fig.~\ref{fig::relic_densities}.

In this regime, the relic abundance constrains the DM coupling $g_\chi$, as seen in the central panel of Fig.~\ref{fig::relic_densities}, with annihilation cross-section $\braket{\sigma v} \propto g_\chi^2 c_Y^2 / m_\chi^2$.  We plot $m_\chi$ against $g_\chi$ in Fig.~\ref{fig::mchi_vs_gchi}; the lower bound is set by the resonance region, and is (unsurprisingly) poorly sampled for low values of $m_\chi$.

To show the impact of allowing DM to be underabundant, we perform a separate scan where we instead employ a Gaussian likelihood for the relic abundance. This can be achieved by instead using the following entry in the \yaml{Rules} section of the \GB input file:

\begin{lstyaml}
  # Choose to implement the relic density likelihood
  # as a detection, not an upper bound
  - capability: lnL_oh2
    function: lnL_oh2_Simple
\end{lstyaml}

We show the $m_\chi$--$m_Y$ plane for this scan in the lower panel of Fig.~\ref{fig::mchi_vs_mY}. For a given point in the $m_\chi$--$m_Y$ plane, the couplings $g_\chi$ and $c_Y$ must be correctly tuned to fit the relic density requirement: clearly, the scanner struggles to find such points compared to when DM can be underabundant. Notably, the sampler struggles to find the very fine-tuned points on resonance when $\overline{t}t$ is not kinematically accessible.

In the lower panel of Fig.~\ref{fig::mchi_vs_gchi} we show the $m_\chi$--$g_\chi$ plane when requiring that $\chi$ fits the observed relic abundance, coloured by $\frac{m_Y}{m_\chi}$. There is a well-defined red area scattered along a straight line with $\frac{m_Y}{m_\chi} < 1$, corresponding to efficient annihilation to $YY$, \ie for $\braket{\sigma v} \propto g_\chi^4/m_\chi^2$.  This is reflected in the lower panel of Fig.~\ref{fig::mchi_vs_mY}: almost all of the valid samples for $m_\chi < m_t$ are in the regime where $m_\chi > m_Y$, \ie above the red dashed line. Here we see that the slope of the line followed by the red area in the lower panel of Fig.~\ref{fig::mchi_vs_gchi} is exactly half that of the lower bound on $g_\chi$, due to the fact that the latter is instead set by resonant annihilation to fermions, which involves one less power of $g_\chi$ in the corresponding matrix element, \ie $\braket{\sigma v} \propto g_\chi^2
c_Y^2 / m_\chi^2$.

Direct detection processes proceed via $t$-channel $Y$ exchange. The functional form of the spin-independent cross-section is \cite[e.g.][]{Abdallah:2015ter}:
\begin{equation}
  \sigma_{\rm SI}^{N} = \frac{\mu_{\chi N}^2 m_N^2}{\pi}\left(\frac{g_\chi c_Y}{v m_Y^2}\right)^2 f_N^2\,,
\end{equation}
where $\mu_{\chi N}$ is the DM-nucleon reduced mass, $N=n,p$, and the form factor
\begin{equation}
f_N = \sum_{q=u,d,s} f^q_{N} + \frac{2}{27}f^G_N\,.
\end{equation}
Here the light-quark form factors are
\begin{align}
  f_p^u &= 0.0233, f_p^d = 0.0343, f_p^s = 0.0458 \\
  f_n^u &= 0.0160, f_n^d = 0.0499, f_n^s = 0.0458,
\end{align}
and the gluon factors $f^G_N = 1-\sum_{q=u,d,s} f^q_{N}$ are
\begin{align}
  f^G_p &= 0.8966, f^G_n = 0.8883.
\end{align}
These follow directly from the values $\sigma_s=43$\,MeV, $\sigma_l=58$\,MeV chosen in the \GB input file presented in Sec.\ \ref{sec:pheno}. Details of the conversion between the two parameterisations, and a discussion of possible values for $\sigma_s$ and $\sigma_l$, can be found in Refs.\ \cite{DarkBit,Cline:2013gha}.

Thus for a given DM mass $m_\chi$, direct detection constrains the parameter combination $g_\chi c_Y/m_Y^2$, rescaled by the DM fraction $f \equiv \Omega_\chi/\Omega_{\mathrm{DM}}$.

\begin{figure}[tbp]
  \centering
  \includegraphics[height=0.84\columnwidth]{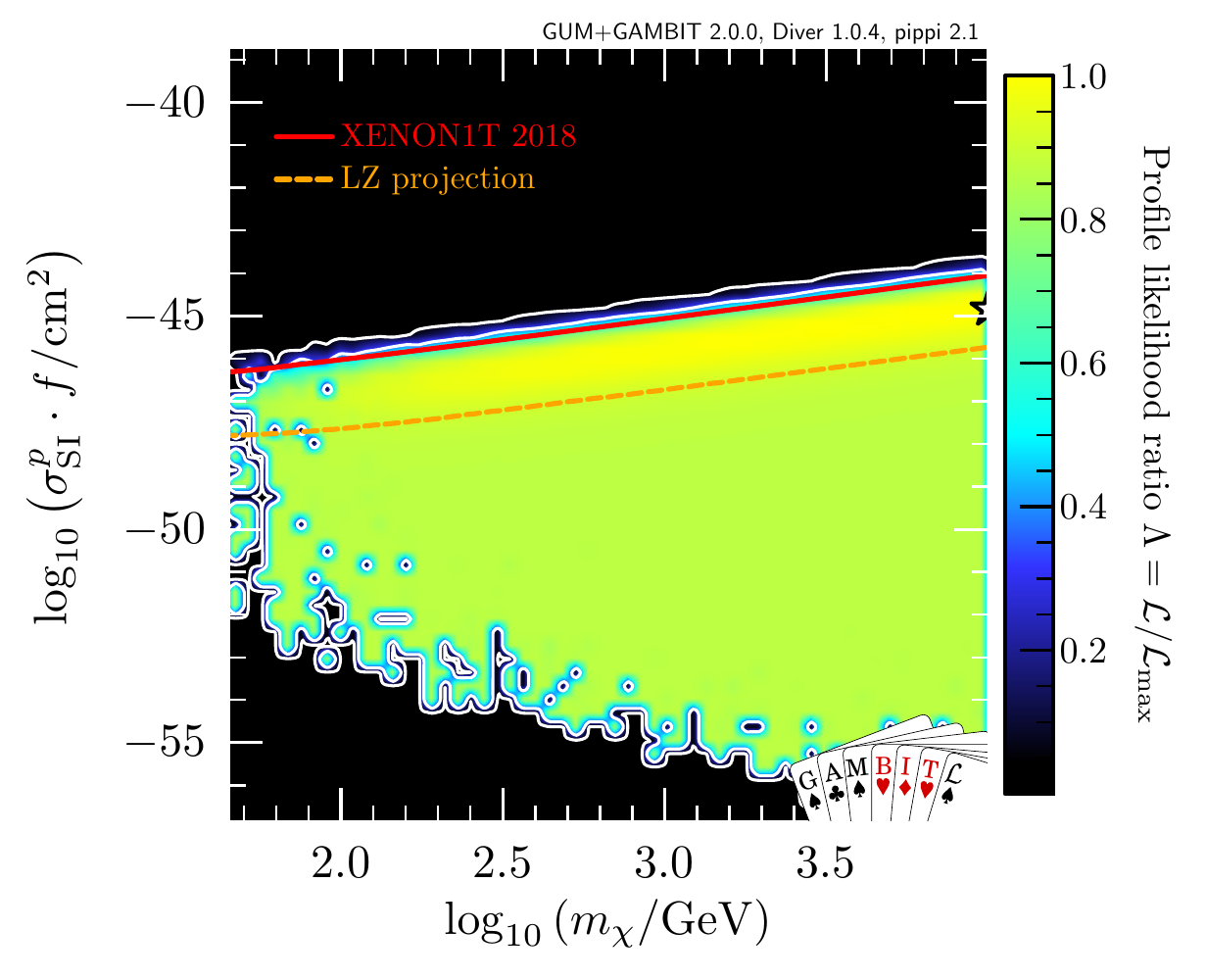}
  \vspace{-2mm}
  \caption{Profile likelihood in the $\sigma_\mathrm{SI}^p$--$m_\chi$ plane. The solid red line shows the exclusion from XENON1T~\cite{Aprile:2018dbl} and the dotted orange line shows the projection from LZ~\cite{LZ}. The spin-independent scattering cross-section of dark matter with protons $\sigma_\mathrm{SI}^p$ is rescaled by the fraction of predicted relic abundance $f\equiv\Omega_\chi/\Omega_{\mathrm{DM}}$. Contour lines show the $1$ and $2\sigma$ confidence regions. The white star shows the best-fit point.}
  \label{fig::cross_sec}
\end{figure}

Fig.~\ref{fig::cross_sec} shows the spin-independent cross-section on protons as a function of the DM mass. As it is possible for $\chi$ to be underabundant for all masses, it is easy to evade direct detection limits by simply tuning the couplings. We also plot the projection from LZ~\cite{LZ}, which shows the significant effect that future direct detection experiments can have on the parameter space of this model, including the ability to probe the current best-fit point.

The best-fit region in Fig.~\ref{fig::cross_sec} lies just below the XENON1T limit: this is due to a small excess (less than $2\sigma$) in the data, which can be explained by this model.  This excess is discussed in more detail in a \GB study of scalar singlet DM~\cite{SSDM2}.

\begin{figure}[tbp]
  \centering
  \includegraphics[height=0.84\columnwidth]{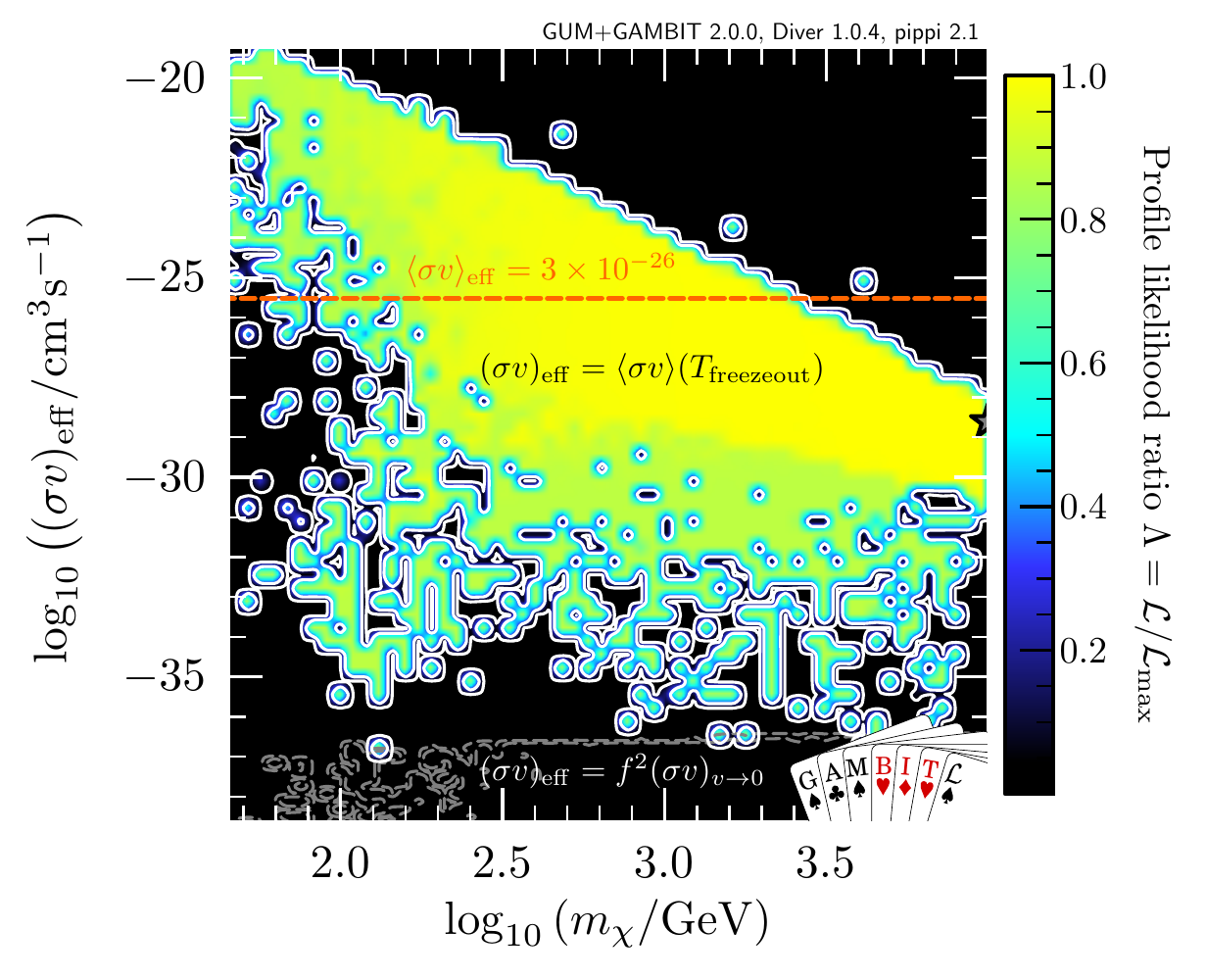}
  \vspace{-2mm}
  \caption{Profile likelihoods for the effective annihilation cross-section as a function of dark matter mass, at different epochs in the evolution of the Universe. The dashed orange line signifies the canonical thermal cross-section $\langle \sigma v \rangle = 3 \times 10^{-26} \text{cm}^3\text{s}^{-1}$. White contours with coloured shading correspond to the thermal average cross-section at dark matter freezeout.  The contours show the $1$ and $2\sigma$ confidence regions, and the white star the best-fit point.  For comparison, in grey contours we show the $1$ and $2\sigma$ confidence regions for the annihilation cross-section in the $v\to0$ limit, rescaled by the square of the fraction of the predicted relic abundance, $f\equiv\Omega_\chi/\Omega_{\mathrm{DM}}$. This is the effective annihilation cross-section that enters indirect detection rates in the late Universe.
  \label{fig::sigmav}}
\end{figure}

Note that for all annihilation channels, the annihilation cross-section is proportional to the square of the relative velocity of DM particles in the direction perpendicular to the momentum transfer, i.e. $\langle \sigma v \rangle \varpropto v_{\perp}^2$.  This means that annihilation is velocity suppressed, especially in the late Universe where $v_{\perp} \sim 0$.  As annihilation processes are also suppressed by the square of the DM fraction $f$, indirect detection signals therefore do not contribute significantly to the likelihood function.  The velocity dependence of the cross-section is fully taken into account by \mo in computing the relic density, however, so the thermally averaged value at freezeout is much larger than the late-time value.  We show the thermally-averaged value at freezeout in Fig.~\ref{fig::sigmav}, which, as expected, overlaps the canonical thermal value $\langle \sigma v \rangle = 3 \times 10^{-26} \text{cm}^3\text{s}^{-1}$.  For comparison, in grey contours we also plot $f^2(\sigma v)_{\rm v \rightarrow 0}$, the effective cross-section for indirect detection.  In this case, all parameter combinations give cross-sections several orders of magnitude below the canonical thermal value, heavily supressing all possible indirect detection signals.

If we wish to remove the model from \GB, we simply run the command
\begin{lstterm}
./gum -r @mug\_files/MDMSM.mug@
\end{lstterm}
and the \GB source reverts to its state prior to the addition of the model.

\section{Summary}\label{sec:summary}

The standard chain for a theorist to test a BSM theory against data has been greatly optimised, and largely automated in recently years, with the development of Lagrangian-level tools such as \fr and \sarah. On the phenomenological side, \GB has been designed as a modular global fitting suite for extensive studies of BSM physics. \gum adds the final major missing piece to the automation procedure.  By providing an interface between \GB, \sarah and \fr, it makes global fits directly from Lagrangians possible for the first time.  This will make the process of performing statistically rigorous and comprehensive phenomenological physics studies far easier than in the past.

We have shown that \gum produces sensible results for a simplified model, in good agreement with previous results found in the literature.  This is based on a scan that can be performed on a personal computer in a reasonable time frame.

The modular nature of \gum means that extension is straightforward.  Since the first version of this paper appeared, \gum has already been extended in \gambit 2.1 to include a four-fermion EFT plugin connecting \fr and \CH \cite{DMEFT}.  Other extensions planned include computation of modifications to SM precision observables and decays, multi-component and co-annihilating dark matter models, and interfacing to the \GB flavour physics module \flavbit via \textsf{FlavorKit}.  We also plan to add new interfaces to public codes not included in this release, including to spectrum generators and decay calculators associated with \FS, and to the dark matter package \mdm. We will also update supported backends to the latest versions, in particular \mo 5 \cite{Belanger:2018ccd}, \higgsbounds 5 \cite{Bechtle:2020pkv}, \higgssignals 2 \cite{Bechtle:2020uwn} and \pythia 8.3.


\begin{acknowledgements}
We thank the rest of the GAMBIT community, in particular Felix Kahlhoefer, for many helpful discussions, and for helping to develop and test \GB over a period of many years. We also acknowledge PRACE for awarding us access to Marconi at CINECA, Italy, and Joliot-Curie at CEA, France. This project was also undertaken with the assistance of resources and services from the National Computational Infrastructure, which is supported by the Australian Government. We thank Astronomy Australia Limited for financial support of computing resources. Computations were also performed on resources provided by UNINETT Sigma2, the National Infrastructure for High Performance Computing and Data Storage in Norway, under project nn9284k. TEG is supported by DFG Emmy Noether Grant No.~KA 4662/1-1. PS is supported by the Australian Research Council (ARC) under grant FT190100814. JECM is supported by the Carl Trygger Foundation through grant no.\ CTS 17:139. JJR acknowledges support by Katherine Freese through a grant from the Swedish Research Council (Contract No. 638-2013-8993). PA, CB and TEG are supported by the ARC under grant DP180102209.  The work of PA was also supported by the Australian Research Council Future Fellowship grant FT160100274.
\end{acknowledgements}

\appendix
\setcounter{table}{0}
\renewcommand\thetable{A\arabic{table}}
\renewcommand\thetocsection{\Alph{section}}

\section{Collider constraints on the Majorana DM simplified model with scalar mediator}
\label{app:collider_validation}

We argued in Sec.\ \ref{sec:example} that the collider constraints are expected to be subleading for the MDMSM. To justify and clarify that argument, and to demonstrate \gum's ability to generate code for collider simulations, we here investigate the likelihood contribution from LHC searches.

It has been demonstrated  that monojet searches are not necessarily the most constraining searches for the MDMSM~\cite{Buckley:2014fba, ATLAS:2021shl}. In fact, given the large Yukawa couplings the tree level production of top quark pairs together with the mediator $Y$, despite the large final state masses, should be the most sensitive final state at the 13 TeV LHC.
To investigate the constraints from this process, we select a 139 fb$^{-1}$ ATLAS search for final states with two leptons, jets and missing momentum, which is targeted to this specific final state~\cite{2102.01444}.

The computational requirement of a \gambit scan increases significantly when full collider simulations with \colliderbit are included. For this example scan we therefore only vary the mass $m_Y$ of the mediator particle. The simulations are performed using the \gum-generated \pythia interface, as described in sections \ref{sec:colliderbit} and \ref{sec:pythia}. For each parameter point in the scan we generate 12 million \pythia events. The events are then passed through fast detector simulation in \colliderbit and selection cuts emulating the ATLAS search. This search targets events with two opposite-charge leptons, jets and missing transverse momentum. No large excesses are observed in this search; across all signal regions the observed event counts agree with the Standard Model expectations to around the 2$\sigma$ level.

The ATLAS analysis defines both exclusive and inclusive signal regions based on the `stransverse mass' kinematic variable and the signal lepton flavours. For our scan we consider the exclusive signal regions. There is no publicly available full likelihood function for this analysis, nor any data on correlations, and we therefore take the conservative approach of only using the likelihood contribution from the single signal region with the best expected sensitivity at each point in our scan.

\begin{figure}[tbp]
    \vspace{-2mm}
    \centering
    \includegraphics[height=0.84\columnwidth]{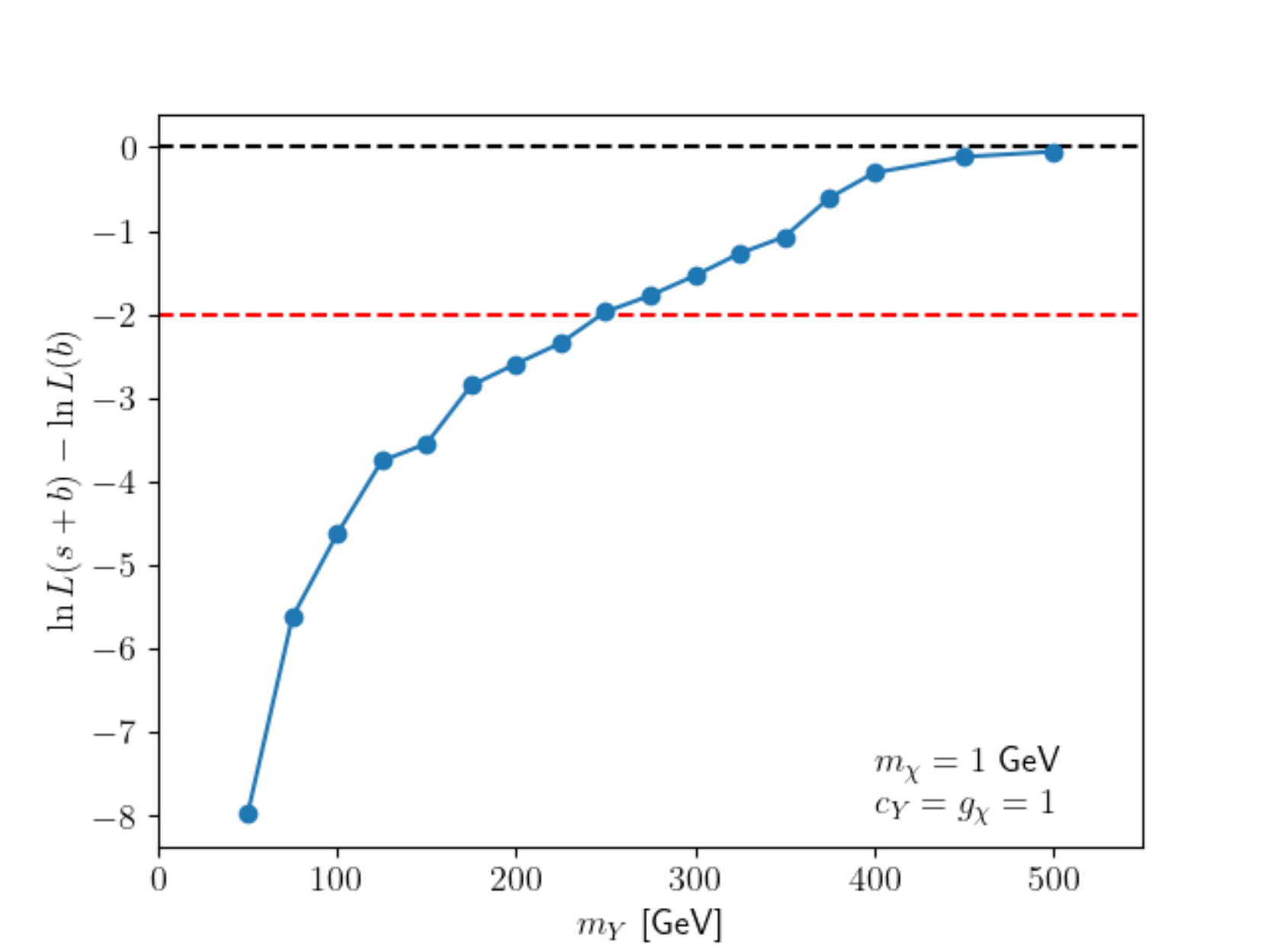}
    \caption{The log-likelihood contribution $\Delta \ln L = \ln L(s+b) - \ln L(b)$ from a simulation of the ATLAS search in Ref.\ \cite{2102.01444} in a scan of the mediator mass $m_Y$ in the MDMSM. The mediator coupling to SM particles ($c_Y$) and to DM ($g_{\chi}$) are set to $c_{Y} = g_{\chi} = 1$, and the DM mass is fixed at $m_{\chi} = 1$\, GeV. The dashed black line denotes $\Delta \ln L = 0$, i.e.\ the limit where the signal plus background prediction fits the data equally as well as the background-only prediction. The dashed red line at $\Delta \ln L = -2$ shows the $\Delta \ln L$ limit corresponding to a $2\sigma$ confidence interval on $m_Y$. The most sensitive signal region differs for the $m_Y = 50$ GeV point, but is otherwise consistent across the scan.}
    \label{fig:MDMSMCollider}
\end{figure}

Figure \ref{fig:MDMSMCollider} shows the resulting ATLAS likelihood function in our scan of the mediator mass $m_Y$, with the other model parameters set to $m_{\chi} = 1$\,GeV and $c_Y = g_{\chi} = 1$. Following the standard approach in \colliderbit, we show the log-likelihood difference $\Delta \ln L = \ln L(s+b) - \ln L(b)$, where $L(s+b)$ denotes the likelihood when the predicted DM signal ($s$) is added on top of the SM background expectation ($b$), and $L(b)$ is the likelihood for the background-only prediction. Further details on the likelihood evaluation in \colliderbit are given in Ref.\ \cite{ColliderBit}. The red dashed line shows $\Delta \ln L = -2$, corresponding to the $\Delta \ln L$ limit for the approximate $2\sigma$ confidence interval on $m_Y$. For $c_Y = g_{\chi} = 1$ and $m_{\chi} = 1$\,GeV, mediator masses below $\sim250$\,GeV are disfavoured at the $2\sigma$ level, in agreement with the constraint that the ATLAS analysis obtains on the mediator mass in a similar BSM scenario. When compared to the $[45,\, 10^4]$\,GeV scan range for $m_Y$ in the MDMSM scan in Sec.\ \ref{sec:example}, and further taking into account that the four-dimensional scan in Sec.\ \ref{sec:example} allowed couplings as small as $10^{-4}$, we see that the collider likelihood would have had a minimal impact on the profile likelihood results of that scan.

\section{New backend interfaces}

\subsection{\CH} \label{app:calchep}

\CH provides squared matrix elements for a given process at tree level. The \GB interface to \CH contains two simple convenience functions, \cpp{CH_Decay_Width} and \cpp{CH_Sigma_V}, which apply the correct kinematics to convert a matrix element into a $1\rightarrow2$ decay width, or a $2\rightarrow2$ DM annihilation cross-section.

The function \cpp{CH_Decay_Width} is used by \decaybit to add a new \cpp{Entry} to its \cpp{DecayTable}. To obtain the decay width, one simply passes the name of the model and the decaying particle as they are known internally in \CH, along with a \cpp{std::vector<std::string>} containing the names of the decay products (also as known to \CH).  Note that at present only two-body final states are allowed by \CH, but the interface generalises nearly automatically to higher-multiplicity final states.

The function \cpp{CH_Sigma_V} returns the product $\sigma v_{\rm lab}$ for DM annihilation $\chi + \overline{\chi} \rightarrow A + B$. It does not support co-annihilations.
This function is used by the \darkbit Process Catalogue. The arguments for \cpp{CH_Sigma_V} are identical to \cpp{CH_Decay_Width}, except that the in states must be a \cpp{std::vector<std::string>} containing the DM candidate and its conjugate. The function also requires the relative velocity in the centre-of-mass frame \cpp{double} \cpp{v_rel} (in units of $c$), and the \cpp{DecayTable}, to pass updated mediator widths to \CH.

For matrix elements with numerical instabilities for zero relative velocity, we compute the cross-section at a reference velocity of $v_{\rm lab} = 1 \times 10^{-6}$.

\subsection{\sarah-\spheno} \label{app:spheno}

\spheno is a spectrum generator capable of providing one-loop mass spectra as well as decay rates at tree and loop level. \gambit has included a frontend interface to the release version of \spheno \textsf{3.3.8} since \GB \textsf{1.0}, and to \textsf{4.0.3} since \gambit \textsf{1.5}. Details about the interface can be found in Appendix B of \cite{SDPBit}. There are important differences between the frontend interfaces to the release version of \spheno and to the \sarah-generated version (which we refer to as \sarah-\spheno).  We give details of these differences below.

\sarah generates the \Fortran \spheno files to compute the spectrum and decays for a given model. These differ from the out-of-the-box \spheno, which only works with various versions of the MSSM. After generating these files with \sarah, \gum moves them to the main \GB directory, to be combined with the downloaded version of \spheno at build time.

In order to improve the usability of \sarah-\spheno in \GB, we have patched two variables into the \Fortran code. The first, \fortran{ErrorHandler_cptr}, is a pointer to a \cpp{void} function that returns control to \GB after a call to the \spheno subroutine \fortran{TerminateProgram}. This prevents \GB being terminated when \spheno fails. Instead, it raises an \cpp{invalid_point} exception, and carries on. The second new variable is \fortran{SilenceOutput}, which provides a \GB input option that allows the user to silence all output of \sarah-\spheno to \term{stdout}.  This option defaults to \yaml{false}.
\begin{lstyaml}
Rules:
  - capability: unimproved_@\nm@_spectrum
    function: get_@\nm@_spectrum_SPheno
    options:
      SilenceOutput: true #default: false
\end{lstyaml}

The interface to the spectrum computation from \sarah-\spheno remains fairly similar to that described for the release version of \spheno in Ref.\ \cite{SDPBit}. Some variables and functions have changed names and library symbols.  The computations have been re-ordered slightly, but otherwise remain unperturbed.

The major change to the spectrum is the computation of mass uncertainties, while the previous \spheno interface merely applied a universal uncertainty to all masses.  These uncertainties are computed by \spheno for all spectrum masses and added to the \GB spectrum object if requested, using the option \fortran{GetMassUncertainty}.  This option defaults to \yaml{false}.
\begin{lstyaml}
Rules:
  - capability: unimproved_@\nm@_spectrum
    function: get_@\nm@_spectrum_SPheno
    options:
      GetMassUncertainty: true #default: false
\end{lstyaml}
Setting \yaml{GetMassUncertainty: true} causes the mass uncertainties to be added to the spectrum in the SLHA block \term{DMASS}.

The most significant difference between the frontend interface to \sarah-\spheno compared to the release version of \spheno is that the former includes computation of decays. The backend convenience function \cpp{run_SPheno_decays} provides a new capability \cpp{SARAHSPheno_}\nm\cpp{_decays}, which maps the decay widths computed by \spheno into a \GB\ \cpp{DecayTable}. Internally, this backend function fills the \sarah-\spheno internal variables with the spectrum details and computes all the branching fractions using the \sarah-\spheno function \fortran{CalculateBR}. Note that the branching fractions for charged processes are rescaled within the frontend interface, as they are double-counted in \spheno so must be rescaled by a factor of $1/2$. Various \GB input options are added for the computation of the decays, the most notable of which are \begin{itemize}
\item \fortran{OneLoopDecays}, which switches on the alternate computation of full one-loop decays,
\item \fortran{MinWidth}, which specifies the minimum width value in GeV for a decay to be added to the table, and
\end{itemize}
These can be given as options of the \cpp{decay_rates} capability as
\begin{lstyaml}
  - capability: decay_rates
    function: all_@\nm@_decays_from_SPheno
    options:
      OneLoopDecays: false@\ \ \ @#default: false
      MinWidth: 1e-10@\ \ \ \ \ \ \ \ @#default: 1e-30
\end{lstyaml}

Lastly, the new \sarah-\spheno interface provides information about Higgs couplings via the backend convenience function \cpp{get_HiggsCouplingsTable}, which provides the capability \cpp{SARAHSPheno_}\nm\cpp{_HiggsCouplingsTable}. This function simply fills a \GB\ \cpp{HiggsCouplingsTable} object from various internal variables in \sarah-\spheno.

\subsection{\veva (C\xx)} \label{app:vevacious}

{
  \renewcommand{\arraystretch}{1.6}
  \begin{table*}[t]
    \centering
    \begin{tabular}{p{6cm} p{10cm}}
    \toprule
    Option                                        & Utility \& Default Value\\ \midrule
    \yaml{phc_random_seed}                         & Set the seed for the routines in \textsf{PHC}. Default: generated by \GB. \\
    \yaml{minuit_strategy}                         &  Select the strategy for \textsf{MINUIT} when minimising the one-loop potential and finding the optimal tunneling path. Higher values mean more function calls and more accuracy. Possible values: \yamlvalue{0}, \yamlvalue{1}, \yamlvalue{2}. Default: \yamlvalue{0}. \\
    \yaml{potential_type}                          & Selects the potential class within \veva. Current available options are (default) \yamlvalue{FixedScaleOneLoopPotential}, \yamlvalue{PotentialFromPolynomialWithMasses}. \\ 
    \yaml{homotopy_backend}                        & Selects choice of software to perform the homotopy continuation, choice of \yamlvalue{hom4ps} and \yamlvalue{phc}. Default: \yamlvalue{hom4ps}. \\
    \yaml{path_finding_timeout}                    & Maximum time spent trying to find the optimal tunneling path, in seconds. Default: \yamlvalue{3600}. \\
    \yaml{survival_probability_threshold}          & The threshold probability for which \veva stops trying to find a lower bounce action, as a fraction of age of the Universe. Default: \yamlvalue{0.01}. \\
    \yaml{radial_resolution_undershoot_overshoot}  & Sets the choice of length scale resolution for the numerical integration of bounce solutions. Default: \yamlvalue{0.1}. \\
    \yaml{PathResolution}                          & Number of equally-spaced nodes along candidate paths in field space that are moved for finding the optimal tunneling path. Default: \yamlvalue{1000}. \\
    \bottomrule
    \end{tabular}
    \caption{Table of the \cpp{runOptions} available to the module function \cpp{initialize_vevacious}, which is used to pass runtime options to \veva.}
    \label{tab:vevaopts}
  \end{table*}
}

\veva computes the stability of the EWSB vacuum in BSM theories when deeper vacua exist. It does so by first finding all minima of the tree-level potential, calculating one-loop and thermal corrections, and computing the likelihood for our vacuum not to have decayed before the present epoch. \veva has been recently rewritten in C\xx and without dependence on external tools for the tunnelling calculation. This is the version that \gum uses. The \GB interface for \veva (C\xx) is described in detail in \cite{VS_GUT}, so we will only summarise it briefly here.

Out of the box, \veva simply requires an SLHA2 file as input. To avoid file operations, from \GB the spectrum object is passed to the central \veva object by the capability \cpp{pass_spectrum_to_vevacious}. It is this capability in \specbit for which \gum can write a new module function for each model.

Besides the spectrum, \veva requires other information to initialise prior to running the main routines. The various operational options for \veva are set via the module function \cpp{initialize_vevacious} within \specbit. These are described in Table~\ref{tab:vevaopts}. The specific minima to which \veva must compute the tunnelling probability are extracted from the \GB input file by the \cpp{panic_vacua} capability, and the specific tunnelling strategy by \cpp{tunnelling_strategy}. The capability \cpp{compare_panic_vacua} checks whether the minima requested are identical.

The main \veva computations are performed using the method \cpp{RunPoint} from the \cpp{VevaciousPlusPlus} class, native to \veva. \GB has access to this class dynamically via the class structure generated by \BOSS and calls this method in the capability \cpp{check_vacuum_stability_vevacious}.

The likelihood of tunnelling to any minimum is provided by the capability \cpp{VS_likelihood}. The selection of which minimum to which to compute the transition and the tunnelling strategy is done by using its \yaml{sub_capabilities}, as described in section \ref{sec:veva_desc}. Lastly, the details of the tunnelling computations by \veva can be extracted as a map using the capability \cpp{VS_results}.


\bibliography{R2}


\end{document}